\documentclass[%
 aip,
 amsmath,amssymb,
 reprint,%
 floatfix
]{revtex4-1}

\usepackage{graphicx}
\usepackage{dcolumn}
\usepackage{bm}

\usepackage[utf8]{inputenc}
\usepackage[T1]{fontenc}
\usepackage{mathptmx}
\usepackage{etoolbox}

\makeatletter
\def\@email#1#2{%
 \endgroup
 \patchcmd{\titleblock@produce}
  {\frontmatter@RRAPformat}
  {\frontmatter@RRAPformat{\produce@RRAP{*#1\href{mailto:#2}{#2}}}\frontmatter@RRAPformat}
  {}{}
}%
\makeatother
\begin{document}

\preprint{AIP/123-QED}

\title[Fully-Periodic CPM MD]
  {Fully Periodic, Computationally Efficient Constant Potential Molecular Dynamics Simulations of Ionic Liquid Supercapacitors}

\author{Shern R. Tee}
\affiliation{Australian Institute of Bioengineering and Nanotechnology, The University of Queensland, Brisbane, QLD, 4072, Australia}
\email{s.tee@uq.edu.au}
\author{Debra J. Searles}
\affiliation{Australian Institute of Bioengineering and Nanotechnology, The University of Queensland, Brisbane, QLD, 4072, Australia}
\affiliation{School of Chemistry and Molecular Biosciences, The University of Queensland, Brisbane, QLD, 4072, Australia}
\email{d.bernhardt@uq.edu.au}

\date{\today}

\begin{abstract}
Molecular dynamics (MD) simulations of complex electrochemical systems, such as ionic liquid supercapacitors, are increasingly including the constant potential method (CPM) to model conductive electrodes at specified potential difference, but the inclusion of CPM can be computationally expensive. We demonstrate the computational savings available in CPM MD simulations of ionic liquid supercapacitors when the usual non-periodic slab geometry is replaced with fully periodic boundary conditions. We show how a doubled cell approach, previously used in non-CPM MD simulations of charged interfaces, can be used to enable fully periodic CPM MD simulations. Using either a doubled cell approach, or a finite field approach previously reported by others, fully periodic CPM MD simulations produce comparable results to the traditional slab geometry simulations with a nearly double speed-up in computational time. Indeed, these savings can offset the additional cost of the CPM algorithm, resulting in periodic CPM MD simulations that are faster than the non-periodic, fixed-charge equivalent simulations for the ionic liquid supercapacitors studied here.
\end{abstract}

\maketitle

\section{Introduction}

A better understanding of the electrode-electrolyte interface is crucial to further progress in energy storage, electrocatalysis, and other electrochemical applications, many of which are vital for meeting the historic challenge of climate change. Room-temperature ionic liquids (RTILs) are especially promising as electrochemical solvents due to their high ionic conductivity and thermal stability, wide electrochemical window and liquid range \cite{Liu2010IonicElectrochemistry,Lian2019HuntingWindows} and extensive tunability \cite{Hayes2015StructureLiquids}. However, modelling the interface between RTILs and electrodes is especially challenging due to the high ionic concentrations of RTILs, for which models beyond mean-field theory are required to take into account dynamic ionic correlations, both within the RTIL and with the electrode\cite{Kornyshev2007Double-layerChange}.

Classical molecular dynamics (MD) is an important technique for studying the interface between electrodes and RTILs, as well as other concentrated electrolytes, to provide insights beyond the current theoretical models. More detailed quantum mechanical methods, such as ab-initio MD, can provide more fundamental models, but cannot reach the nano- to microsecond timescales required for observing ionic layer rearrangement and other slow electrolyte phenomena, which is routinely achievable in classical MD simulations. Therefore, improving MD simulations of RTILs is an active area of research, with many recent promising calibrations of polarizable \cite{Bedrov2019MolecularFields}, atomistic \cite{Doherty2017RevisitingSimulations} and coarse-grained \cite{Roy2010AnModel,Fajardo2020MolecularForce-field} force-fields for RTILs.

As electrolyte force fields become more sophisticated and realistic, more effort should be invested at the same time into modelling conductive electrodes with more realistic dynamics. Most MD simulations consider an oversimplified model of the electrode, where charge is simply uniformly distributed across the electrode surface and remains fixed over time. This fixed-charge method (FCM) does not maintain a constant potential across the electrode surface, and thus does not accurately model a conductive electrode. In addition, it clearly omits phenomena where the surface charge changes over time, the most prominent being capacitor charging and discharging.

By contrast, the constant potential method (CPM) \cite{Siepmann1995InfluenceSystems,Reed2007ElectrochemicalElectrode,Gingrich2010OnSurfaces,Tazi2010Potential-inducedInterface} explicitly includes charge redistribution steps to better model conductive electrodes. CPM MD generally provides better accuracy for simulations of electrochemical interfaces  \cite{Wang2014EvaluationCapacitors,Haskins2016EvaluationLayers}, and is particularly important for understanding non-planar electrodes \cite{Xing2013OnPores,Merlet2013SimulatingSurfaces,Vatamanu2017OnLayers}. Capturing the dynamic local fluctuations in electrode charges is also indispensable when studying dynamical phenomena of electrochemical interfaces, which influence capacitor charging and discharging \cite{Noh2019UnderstandingSimulations,Demir2020InvestigationSimulations}, electrochemical thermodynamics \cite{Merlet2013SimulatingSurfaces}, and electroresponsive tribology \cite{Seidl2021MolecularElectrolytes}. The technical aspects of CPM MD are themselves a burgeoning field of research, with modifications recently proposed to model electrode metallicity in addition to conductivity \cite{Nakano2019ASimulations,Scalfi2020ASimulations}. A recent publication\cite{Ahrens-Iwers2021ConstantMesh} describes independent upgrades to the CPM implementation in the Large-scale Atomic/Molecular Massively Parallel Simulator (LAMMPS) software package\cite{LAMMPS}, including adaptations to mesh-based long ranged electrostatic evaluation, that are complementary to the implementations in this paper\cite{Tee2022Repo,GitHubLink} and raise the possibilities for significant efficiency gains.

Nonetheless, computational cost remains a significant obstacle to more widespread adoption of CPM MD, since every charge update step includes at least one re-evaluation of the system's overall electrostatic energy. In any MD simulation with a large proportion of charged particles, the Fourier space calculation of long-ranged electrostatic interactions are usually the most computationally intensive component. Simulations of systems with mixed periodicity -- such as electrode-electrolyte systems, which are not periodically repeated transverse to the electrodes -- are even more expensive, even considering the ``slab correction'' \cite{Yeh1999EwaldGeometry} techniques often employed.

Recently, two fully periodic approaches for studying electrochemical interfaces have been proposed and explored as fully periodic alternatives to the slab correction: finite field simulations \cite{Dufils2019SimulatingElectrode} and doubled cell simulations \cite{Raiteri2020MolecularInterface}. In this paper, we demonstrate that these approaches make RTIL-electrode simulations significantly more efficient by enabling fully periodic CPM MD simulations and eliminating the need for slab corrections. Full periodicity substantially reduces the computational cost of evaluating long-range electrostatics, to the extent that fully periodic CPM MD simulations can be faster than non-periodic FCM MD simulations run on the same hardware. As such, the increased accuracy afforded by CPM MD simulations can be achieved with little or even no overhead relative to FCM MD, and we recommend their routine use when studying RTIL-electrode interfaces. A copy of the source code used in this paper is available online\cite{Tee2022Repo,GitHubLink}.

\section{Methods}

\subsection{An overview of CPM MD}
\label{ss:basiccpm}

\begin{figure}
    \centering
    \includegraphics[width=0.95\columnwidth]{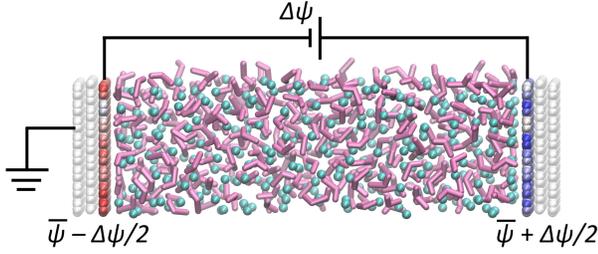}
    \caption{\textbf{Depiction of simulated constant potential supercapacitor}, showing the cations (magenta) and anions (teal) between three-layer graphene electrodes (white). For each electrode, a single proximal layer is charge-updated during CPM MD, leading to either negative (red) or positive (blue) charges induced on each electrode as appropriate. This results in the potential difference $\Delta \psi$ being imposed between the electrodes. The potential offset $\overline{\psi}$ maintains the overall electroneutrality of the system (depicted here as one of the electrodes being grounded).}
    \label{fig:boxpsis}
\end{figure}

Molecular dynamics can provide a detailed model of electric layers in capacitors, using a fixed volume simulation cell containing electrolyte molecules sandwiched between two charged electrodes (Figure \ref{fig:boxpsis}). In CPM MD simulations, the electrode charges $q_i$ are periodically updated to maintain the electrode potentials $\Psi_i$ at their prescribed values \cite{Reed2007ElectrochemicalElectrode,Merlet2013SimulatingSurfaces,Wang2014EvaluationCapacitors}. Here we give a brief overview of the method, as excellent detailed descriptions are available in other recent publications\cite{Scalfi2020ChargeEnsemble,Ahrens-Iwers2021ConstantMesh}.

The potential energy of the MD simulation cell, $U$, is the sum of all non-Coulombic energies $U_{NC}$ and all Coulombic interactions. The Coulombic interactions can further be divided into electrolyte-electrolyte, electrolyte-electrode, and electrode-electrode interactions. Due to the delocalisation of charge on the conductive electrodes, it is usual to represent these as a set of Gaussian functions centred on the atomic sites, whereas the charges on the ions or molecules in the liquid are usually treated as point-charges. Using $i$ to index the electrode atoms and $j$ to index the charged sites of the electrolyte ions and molecules, we can then write $U$ as
\begin{align}
    U = & U_{NC} + \frac{1}{4 \pi \epsilon_0}\Bigg[\sum_{j,j',\,\mathrm{pbc}} \frac{Q_j Q_{j'}}{|\mathbf{R}_j - \mathbf{R}_{j'}|} + \sum_{i,j,\,\mathrm{pbc}} \int \frac{Q_j \rho_i(\mathbf{r})}{|\mathbf{R}_j - \mathbf{r}|} d^3\mathbf{r} \nonumber \\
    &+ \sum_{i,i',\,\mathrm{pbc}} \iint \frac{\rho_i(\mathbf{r}) \rho_{i'}(\mathbf{r}')}{|\mathbf{r} - \mathbf{r}'|} d^3\mathbf{r} \, \, d^3\mathbf{r}'\Bigg]. \label{eqn:U_definition}
\end{align}
Here the subscripts ``pbc'' denote periodic boundary conditions (to be discussed later), $\epsilon_0$ is the permittivity of free space, and $Q_j$ and $\mathbf{R}_j$ are the charge and position of electrolyte charge indexed $j$. The charge density $\rho_i(\mathbf{r})$ associated with electrode particle indexed $i$ is a Gaussian density centered at position $\mathbf{r_i}$,
\begin{align}
    \rho_i(\mathbf{r}) = q_i n_i(\mathbf{r})= q_i \frac{\eta^3}{\pi^{3/2}} \exp \left(-\eta^2 |\mathbf{r}-\mathbf{r}_i|^2 \right) \label{eqn:gauss_definition}
\end{align}
where $q_i$ is the total charge on the electrode atom and $\eta$ (in inverse length units) serves as a width parameter for the Gaussian charges. The use of Gaussian charge densities ensures that the electrode-electrode interactions can later be written as an invertible matrix\cite{Gingrich2010OnSurfaces}.

Writing the vector of the electrode charges $\mathbf{q} \equiv \{q_1, \cdots q_i, \cdots\}$, the simulation box potential energy $U$ (Equation \eqref{eqn:U_definition}) can be written as a quadratic form in $\mathbf{q}$:
\begin{equation}
    U = U_{NC} + U_{elyt} - \mathbf{q}^T \mathbf{b} + \frac{1}{2} \mathbf{q}^T \mathbf{A} \mathbf{q}. \label{eqn:quadform}
\end{equation}
Here $U_{elyt}$ is the sum of the electrolyte-electrolyte Coulombic interactions, the vector $\mathbf{b}$ represents electrolyte-electrode interactions and the matrix $\mathbf{A}$ represents electrode-electrode interactions -- that is, equation \eqref{eqn:quadform} represents the terms (in order) from equation \eqref{eqn:U_definition} as functions of $\mathbf{q}$. The electrostatic potential vector $\boldsymbol{\Psi} \equiv \{\Psi_1, \cdots, \Psi_i, \cdots \}$ is the derivative of the energy with respect to the electrode charges, and therefore
\begin{equation}
    \boldsymbol{\Psi} \equiv \frac{\partial U}{\partial \mathbf{q}^T} = \mathbf{A}\mathbf{q} - \mathbf{b} \label{eqn:psifromu}
\end{equation}
where the second equality follows from equation \eqref{eqn:quadform}. The elements of $\mathbf{b}$ depend on the positions of the electrolyte atoms which will vary with time. However, the electrode atoms are often fixed during a simulation, in which case $\mathbf{A}$ will not vary with time.

In CPM MD, the electrode charges $\mathbf{q}$ are updated so that the electrode potentials are specified. That is, we seek a specific $\mathbf{q^*}$ such that substituting into equation \eqref{eqn:psifromu} gives
\begin{equation}
    \boldsymbol{\Psi} = \overline{\psi} \mathbf{e} + \Delta \psi \mathbf{d} \label{eqn:psiresult}.
\end{equation}
Here $\mathbf{d}$ is an ``indicator'' vector with entries $-1/2$ for elements corresponding to atoms on one electrode and $1/2$ for elements corresponding to atoms on the other,  and $\mathbf{e}$ is a ``sum'' vector with entries 1 for all elements. This general form ensures that atoms of the same electrode have equal potential and there is a potential difference $\Delta \psi$ between the electrodes, while allowing for an overall offset potential $\overline{\psi}$ relative to the potential at infinity. Then $\mathbf{q}^*$ can be directly determined:
\begin{equation}
    \mathbf{q}^* = \mathbf{A}^{-1}( \overline{\psi} \mathbf{e}+\Delta \psi \mathbf{d}+\mathbf{b}). \label{eqn:exact_fixedoffset}
\end{equation}
Other studies, including the prior LAMMPS implementation of CPM MD \cite{Wang2014EvaluationCapacitors} simply adopted $\overline{\psi} = 0$. However, this results in the total charge of the system being non-zero in general; this can severely jeopardize the accuracy of the resulting CPM MD simulation, as recently discussed \cite{Ahrens-Iwers2021ConstantMesh}. Substituting equation  (\ref{eqn:psiresult}) into the electroneutrality constraint $\mathbf{e}^T \mathbf{q}^* = 0$ and solving for $\overline{\psi}$ gives
\begin{equation}
    \overline{\psi} = -\frac{\mathbf{e}^T\mathbf{A}^{-1}(\Delta\psi\mathbf{d}+\mathbf{b})}{\mathbf{e}^T\mathbf{A}^{-1}\mathbf{e}} \label{eqn:charge_neutral}
\end{equation}
in which case the constant potential, electroneutral charge vector $\mathbf{q^*}$ is given by
\begin{align}
    \mathbf{q}^* &= \mathbf{A}^{-1}\left(\Delta \psi \mathbf{d}+\mathbf{b} -  \frac{\mathbf{e}^T\mathbf{A}^{-1}(\Delta\psi\mathbf{d}+\mathbf{b})}{\mathbf{e}^T\mathbf{A}^{-1}\mathbf{e}}\mathbf{e}\right) \nonumber \\ 
    &= \mathbf{O}\mathbf{A}^{-1} \left( \Delta \psi \mathbf{d} + \mathbf{b} \right). \label{eqn:matr}
\end{align}
where the final result is arrived at by defining an ``electroneutrality projector'' matrix,
\begin{equation}\mathbf{O} \equiv \mathbf{I}-(\mathbf{A}^{-1}\mathbf{e}\mathbf{e}^T)/(\mathbf{e}^T\mathbf{A}^{-1}\mathbf{e}).
\label{eqn:odefinition}
\end{equation}
This is the same result recently obtained by considering statistical mechanics on the constant potential ensemble \cite{Scalfi2020ChargeEnsemble}. If the electrode particles used in CPM MD remain stationary, the matrix $\mathbf{O}\mathbf{A}^{-1}$ will be constant and can be precomputed, as its entries only depend on the electrode particle positions. Along these lines, we have updated the previous version of the LAMMPS CPM MD package \cite{Wang2014EvaluationCapacitors} to include the electroneutrality correction, and use the new charge-neutral version \cite{Tee2022Repo,GitHubLink} in the calculations below. 

Then, the main computational burden during each charge update step is to obtain the vector of electrode potentials, $\mathbf{b}$, from the positions of the electrolyte particles. Switching from partially periodic to fully periodic boundary conditions substantially speeds up this step, as we discuss in the next section. 

\subsection{Faster Electrostatic Evaluation Using Fully Periodic Boundary Conditions}
\label{ss:optim}

MD simulations are performed with periodic boundary conditions to enable inferences about macroscopic systems from a nanometer-scale simulation volume. In supercapacitor simulations, the electrodes and electrolyte are usually repeated infinitely parallel to the electrode surface (which we label the $x$ and $y$ axes) but not transverse to the electrodes (which we label the $z$ axis), as depicted in Figure 2(a). Given the slow $1/r$ decay of the Coulombic interaction, direct evaluation of the Coulombic interaction terms in \eqref{eqn:U_definition} is not feasible. Instead Ewald summation can be used, where the Coulomb interaction is truncated in real-space so that it can be treated with a finite cutoff, with the truncated long-range Coulombic interaction being calculated using Fourier transforms in reciprocal space, where it rapidly converges \cite{deLeeuw1980SimulationConstants,Allen1989ComputerLiquids}.

\begin{figure}
    \centering
    \includegraphics[width=\columnwidth]{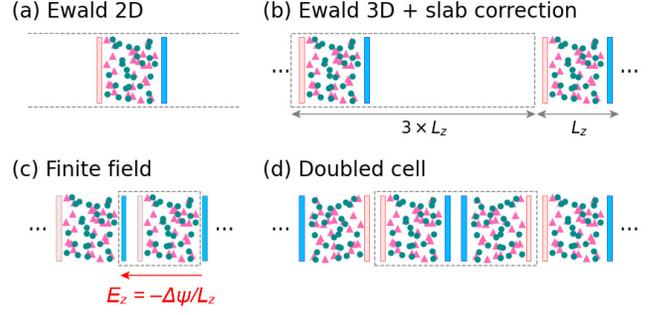}
    \caption{\textbf{Comparison of mixed and fully periodic simulation boxes for CPM MD.} With mixed periodicity, either (a) exact 2D Ewald summation or (b) slab correction must be used for the long-range electrostatics of the simulation box. Full periodicity can be recovered with either (c) the finite field method, that adds an electric field to impose the desired potential difference across the simulation box, or (d) the doubled cell method which combines two cells of reverse polarity to create an overall zero-dipole simulation box.}
    \label{fig:configandsnapshot}
\end{figure}

Although this is straightforward in systems with full, three-dimensional periodicity, the electrolyte-electrode system as described is trickier to handle because of its mixed periodicity. The finite $z$ size of the system changes the associated Fourier sum into a Fourier integral  \cite{Kawata2001RapidPeriodicity,Reed2007ElectrochemicalElectrode}, and this ``exact Ewald 2D'' method is rarely implemented in major MD codes. Instead, the ``slab correction'' is commonly used: as shown in Figure 2(b), $z$-periodicity is imposed with empty space added between repeats of the simulation box, and a charged-sheet approximation is used to remove the residual dipole-dipole interaction between those repeats \cite{Yeh1999EwaldGeometry}.

The slab correction is equivalent to discretizing the exact Ewald 2D method \cite{Brodka2002ElectrostaticSummation}, but still bears increased computational costs: since the simulation box has been expanded in the $z$ direction, more reciprocal vectors are required to reach the same accuracy, resulting in larger Fourier grids and more computational burden for the reciprocal space calculation. This has motivated a search for methods of simulating electrode-electrolyte systems that preserve full periodicity. Two such methods have been recently described in the literature \cite{Dufils2019SimulatingElectrode,Raiteri2020MolecularInterface} and are introduced below. They are applied to the simulation of a model supercapacitor with CPM MD and compared in this paper.

\subsubsection{Finite Field Method}

The first method utilizes an applied ``finite field'' to decouple adjacent simulation boxes \cite{Dufils2019SimulatingElectrode} (Figure 2(c)), motivated by a related approach to modelling polarizability in bulk systems \cite{Zhang2020ModellingDynamics}. In this method, for an intended potential difference $\Delta \psi$, an electric field $E_z = -\Delta \psi / L_z$ is applied across the simulation box.

The simulation box potential energy under a finite field, $U_{ff}$, is obtained by adding a polarization term to the original potential energy $U$ from equation \eqref{eqn:U_definition}:
\begin{equation}
    U_{ff} = U - \frac{\Delta \psi}{L_z}\left(\sum_i q_i z_i+ \sum_j Q_j Z_j\right) \label{eqn:ffield}.
\end{equation}
where $z_i$ and $Z_j$ are the z-positions of the electrode atoms and electrolyte charge sites, respectively. We also define $\mathbf{z}$ as the vector of z-positions of the electrode atoms. (The ``itinerant'' polarization\cite{Zhang2020ModellingDynamics} does not need to be tracked because all particles are bounded between the electrodes.)

The electrode charges $\mathbf{q}$ are then simply obtained by considering the electric field in specification of the potential in equation (\ref{eqn:psiresult}) (i.e. $\boldsymbol{\Psi} = \overline{\psi} \mathbf{e} -\Delta\psi\mathbf{z}/L_z$), and solving for $\mathbf{q}$ with a variable $\overline{\psi}$ and imposing charge neutrality. The field introduces a discontinuity in the potential generated across the box with a value $\Delta\psi$ which is the potential difference between the electrodes. The electrode charges are then,
\begin{equation}
    \mathbf{q}^* = \mathbf{O} \mathbf{A}^{-1}\left(-\frac{\Delta \psi}{L_z}\mathbf{z}+\mathbf{b}\right). \label{eqn:matr_ff}
\end{equation}
which is a simple modification of (\ref{eqn:matr}) and allows an implementation that is similar to the basic CPM MD algorithms. Although the finite field method has very recently been applied to a computational RTIL-electrode interface \cite{Dufils2021ComputationalSimulations}, comparisons have not been made with the slab correction method in terms of either accuracy or computational speed. We document these comparisons and as our first major computational result show that, with properly optimized algorithms, the finite field method is significantly faster thanks to full periodicity in long-range electrostatics evaluations.

\subsubsection{Doubled Cell Method}

Another available method which has been used for other systems, but not for CPM MD, is a ``doubled cell'' approach. In this approach, two sub-cells are built back-to-back with opposing polarities; this yields a unit cell which has zero net dipole, and thus automatically has no dipole-dipole interactions along the $z$ direction. Although the system to be simulated is twice as large, each sub-cell is effectively independent, so that twice as much data can be collected per simulation interval.

This method has previously been applied to polar surfaces as the ``mirrored slab'' method \cite{Croteau2009SimulationConditions,Ren2020EffectsWater}, to capture the dynamics of polar liquids near statically charged surfaces. More recently, this method was applied to simulate the effects of an applied electric field on a liquid-liquid interface  \cite{Raiteri2020MolecularInterface}. However, the same motivation in both cases -- allowing fully periodic electrostatic evaluation for systems with significant overall dipole -- also applies to CPM MD, and we demonstrate as the second major computational result that doubled cell CPM MD also yields accurate results with reduced computational cost relative to slab correction.

The only additional complexity occurs if we require each sub-cell to be independently electroneutral to resemble the single cells. We can accomplish this by modifying equation \eqref{eqn:psiresult} to include two offset potentials:
\begin{equation}
        \boldsymbol{\Psi} = \overline{\psi} \mathbf{e} + \overline{\psi_1} \mathbf{e}_1 + \Delta \psi \mathbf{d} \label{eqn:psi_doublecell}.
\end{equation}
The vectors $\mathbf{e}$, $\mathbf{e_1}$ and $\mathbf{d}$ run over all electrode particles in both sub-cells. As above, the elements of $\mathbf{e}$ are 1 for all electrode particles and the elements of $\mathbf{d}$ are $1/2$ for atoms on both positive electrodes (one for each sub-cell) and $-1/2$ for atoms on the negative electrodes. The new vector $\mathbf{e}_1$ ``selects'' the electrode particles of only one sub-cell -- that is, its elements are 1 for all particles of the electrodes (both positive and negative) in one of the sub-cells , and 0 for all electrode particles in the other sub-cell. When we require that $\mathbf{e}^T \mathbf{q}^* = 0$ and $\mathbf{e}_1^T \mathbf{q}^* = 0$, this ensures that both sub-cells are electroneutral.

We can then write out the two corresponding projection matrices in analogy with equation \eqref{eqn:matr}. Fulfilling the first constraint $\mathbf{e}^T \mathbf{q}^* = 0$, and solving for $\overline{\psi}$, gives us
\begin{equation}
    \mathbf{q}^* = \mathbf{O}\mathbf{A}^{-1}(\overline{\psi_1}\mathbf{e}_1 + \Delta \psi \mathbf{d} + \mathbf{b}) \label{eqn:psi_step_doublecell}
\end{equation}
where $\mathbf{O}$ is defined as above (equation (\ref{eqn:odefinition})). Fulfilling the second constraint, $\mathbf{e}_1^T \mathbf{q}^* = 0$, and solving for $\overline{\psi_1}$ then gives
\begin{equation}
    \mathbf{q}^* = \mathbf{O}_1\mathbf{O}\mathbf{A}^{-1}(\Delta \psi \mathbf{d} + \mathbf{b}) \label{eqn:psiresult_doublecell},
\end{equation}
where
\begin{equation}
    \mathbf{O}_1 \equiv \mathbf{I} - \frac{\mathbf{O}\mathbf{A}^{-1}\mathbf{e}_1\mathbf{e}^T_1}{\mathbf{e}^T_1\mathbf{O}\mathbf{A}^{-1}\mathbf{e}_1}.
\end{equation}

In short, maintaining independent electroneutrality for each cell simply requires pre-calculation of one additional projection matrix then $\mathbf{O}_1\mathbf{O}\mathbf{A}^{-1}$, which does not change in systems with stationary electrode particles, resulting in minimal additional computational burden. Although this does result in a nominal difference in the instantaneous offset potentials \emph{between} the cells, our results show that electrolyte dynamics within one cell is not affected by the configuration in the other. Intuitively, this arises since conductors screen electric fields and therefore the presence of conductive electrodes between the electrolytes of each cell prevents them from interacting, provided the separation of the sub-cells is larger than the cutoff radii for the short-range interactions. This further emphasizes that the offset potential serves only to maintain electroneutrality and does not affect the accuracy of simulation results in any other way.

\subsection{Calculating Potential Profiles In Different Optimized Geometries}
\label{ss:potprof}

The electric potential profile across the simulation box, $\psi(z)$, is a key measurement output for a computational supercapacitor, as it is used to determine the differential capacitance at each electrode, and our third major result is that full periodicity also simplifies the calculation of $\psi(z)$. If the potential profile only varies in $z$, then it can be calculated by obtaining the linear charge density, $\rho(z)$, and solving the 1D Poisson equation:

\begin{equation}
    \frac{d^2}{dz^2} \psi(z) = -\frac{1}{\epsilon_0} \rho(z). \label{eqn:pois}
\end{equation}
subject to the appropriate boundary conditions. For this paper we use a matrix-based finite difference method, which approximates $d^2/dz^2$ as a linear finite difference relation; this method converges well even at low finite-difference orders \cite{Wang2016ElectricCapacitors}. The Poisson equation can then be inverted simply by applying the inverse finite difference matrix to the discretization of the charge density $\rho(z)$.

If electroneutrality is not imposed, the boundary conditions involve setting the potential at each electrode to their CPM pre-specified values (see Supporting Information of  \cite{Demir2020InvestigationSimulations}, for example). If this is not the case, the set potential difference, $\Delta \psi$, and $\overline{\psi}$ can be calculated from equation (\ref{eqn:psiresult}) can be used. Since the both finite field and doubled cell methods have full periodicity, this can be exploited and replace $\Delta \psi$. In the finite field method, the total potential rise $\psi(L_z) - \psi(0)$ is simply the preset potential difference $\Delta \psi$, which enters the boundary conditions as a discontinuity across the $z$-boundary of the unit cell. In the doubled cell method, the (doubled) unit cell is repeated with no further modification and the boundary condition is just continuity across the $z$-boundary, $\psi(L_z) - \psi(0) = 0$. The continuity of the potential is likely to be easier to apply if the electrodes are not planar. Furthermore, our results show that for the special case of planar electrodes, the boundary condition can accommodate the use of CPM MD to only charge or discharge the electrode layers closest to the electrolyte, resulting in further computational savings.

\section{Simulations and Analyses}
\label{sec:simulations}

We demonstrate the slab, finite field, and doubled cell methods with a computational ionic liquid supercapacitor (Figure \ref{fig:boxpsis}). The supercapacitor electrolyte consists of a 10-nm wide block of 1-butyl-3-methylimidazolium hexafluorophosphate (BMim$^+$-PF$_6^-$). The electrolyte is sandwiched between two atomistic graphene electrodes of three layers each, with the interlayer spacing set to 0.335 nm as standard. The ions are simulated using a coarse-grained model that has been tested previously\cite{Roy2010AnModel}. The BMim$^+$ cations and PF$_6^-$ anions are coarse-grained to three particles and one particle per ion respectively, with Lennard-Jones and Coulomb interactions modelled using parameters from the literature\cite{Roy2010AnModel}. In the model, the IL ions are charge-scaled so that the charge on each cation (anion) is +0.78 (-0.78). Literature values for the graphene carbon atom Lennard-Jones parameters \cite{Cole1983TheGraphite} were used and Lennard-Jones parameters for interactions betweeen the carbon and IL atoms were obtained using standard Lorentz-Berthelot rules. The supercapacitor was simulated over a set of potential differences from 0.0 to 2.5 V, for 30 ns per run at each potential difference, and each set of runs was repeated three times for each method from statistically different initial configurations.

During production runs, only the first layer of each electrode closest to the electrolyte was ``charged'' with CPM MD, with the next two layers contributing only non-Coulombic interactions, based on previous studies finding that charge is predominantly induced on the first layer \cite{Wang2014EvaluationCapacitors}, and consistent with the charge distribution of a conductor. Snapshots from each run were separately post-processed to determine the charges that would have been obtained with the constant potential applied to all three layers. The resulting charge distribution was found to confirm the findings of the previous studies, as we discuss later, validating the choice to charge only one layer during production for significant computational savings. In subsequent discussion, these differing configurations of electrode charges are referred to as ``single-layer'' and ``three-layer'' charges respectively.

From each run, the transverse charge profile across the cell was obtained as an equilibrium average and the potential profile calculated using the finite-difference method discussed earlier. The potential on each electrode was subsequently obtained by comparison to the bulk potential, and charge-potential curves were then used to obtain the single-electrode differential capacitance by spline fitting. Further details for the simulations and analyses are given in the appendix.

\section{Results}

\subsection{Charging and Steady State Properties}

\begin{figure}
    \centering
    \includegraphics[width=\columnwidth]{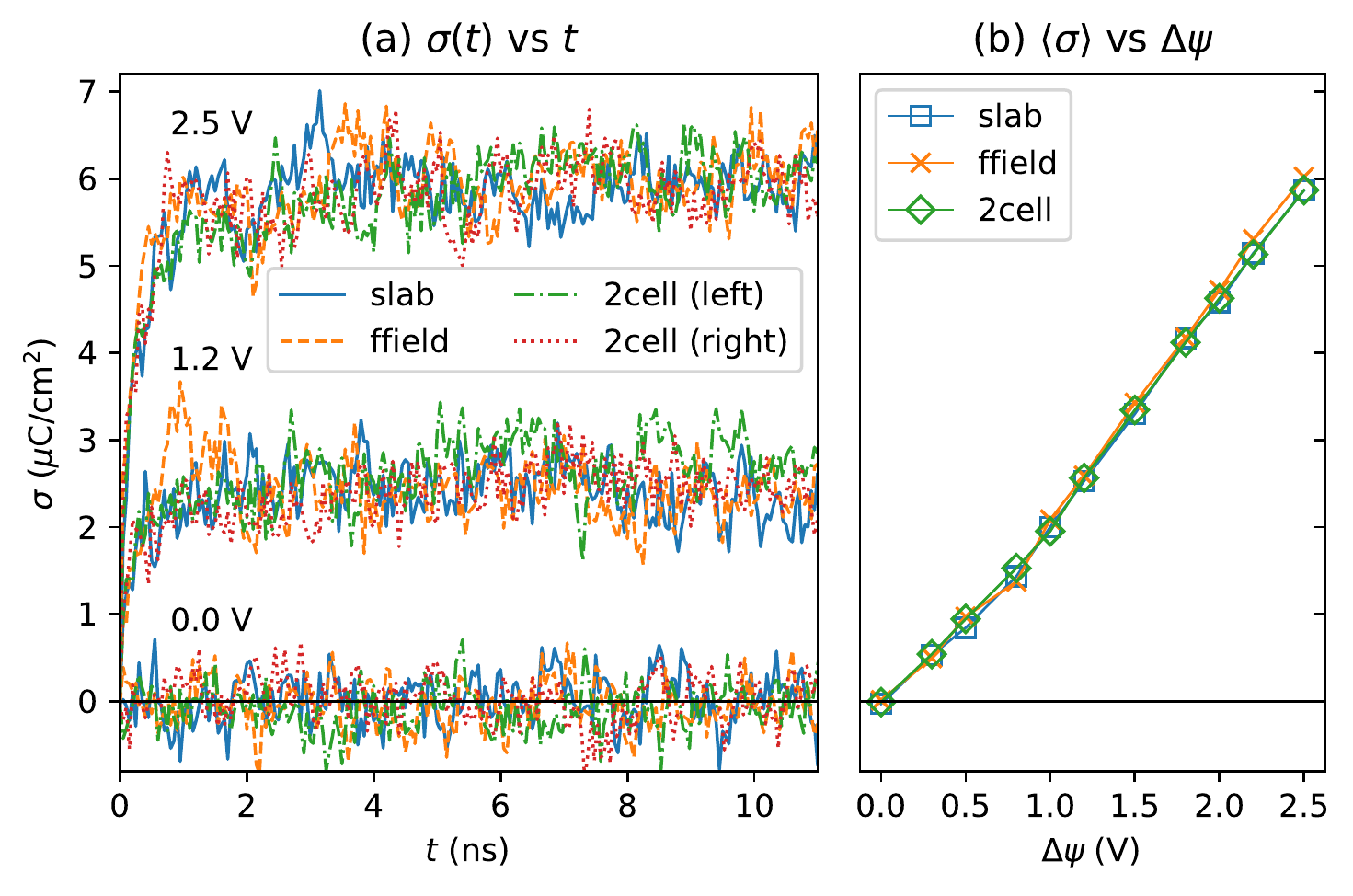}
    \caption{(a) Charging curves of electrode charge against time show that slab, finite field, and doubled-cell methods return statistically similar instantaneous results, as displayed for 0.0, 1.2, or 2.5 V potential difference. (b) The equilibrium surface charge density, averaged over the final 25 ns of each trajectory, as plotted against imposed potential difference $\Delta \psi$ shows that slab, finite field, and doubled-cell methods also return statistically similar ensemble results. Standard error of mean charge (as calculated from the averages of 5 ns trajectory blocks) are smaller than the symbol size. Results from both sub-cells were averaged for the doubled cell data points.\label{fig:stv}}
    \label{fig:my_label}
\end{figure}

Figure \ref{fig:stv}(a) shows typical traces of the surface charge density, $\sigma$, against simulation time, at 0.0, 1.2 and 2.5 V for the first 15 ns of the 30 ns trajectories. These graphs (as well as for other potential differences, supplied in the SI Fig A) show a non-zero charging time characteristic of the CPM MD simulation, allowing equilibration to be visually estimated. Equilibrium charges are attained within a few nanoseconds, which is a typical timescale for coarse-grained simulations, and so the final 25 ns of each trajectory is taken as the equilibrium portion for further analysis.

The simulations using the slab, finite field, and doubled cell geometries return largely identical results, whether from the charging curves or from obtaining the equilibrium average charges as a function of potential difference (Fig 1(b)). To further validate the doubled cell method, we analysed both  long term charges and short term dynamics of doubled cell trajectories.

\subsection{Validating the Doubled Cell Method}
\label{ss:2cell}

\begin{figure}
    \centering
    \includegraphics[width=0.8\columnwidth]{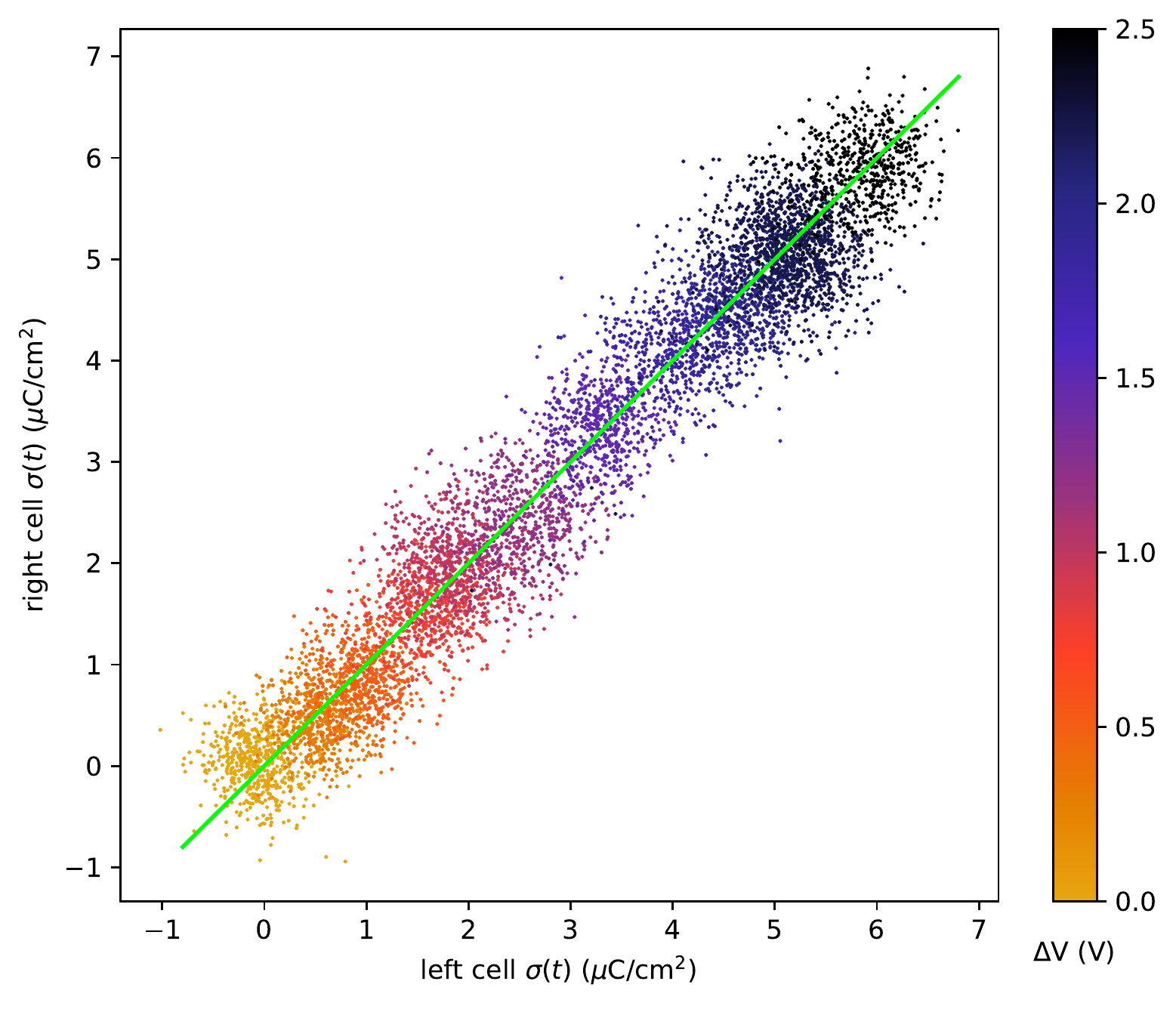}
    \caption{Scatter plot of instantaneous electrode charge densities in doubled cell CPM MD, after equilibration. Left and right cells have identical averages at all potential differences (denoted by point color), as highlighted by the green line (left $\sigma(t)$ = right $\sigma(t)$), but are not noticeably correlated.\label{fig:cellslr}}
\end{figure}

In order for the doubled cell method to be computationally efficient, both cells must return independent trajectories so that the doubled system size truly gives twice as much data. As seen in Figure \ref{fig:cellslr}, the instantaneous electrode charges are indeed uncorrelated between different cells. 

\begin{figure}
    \centering
    \includegraphics[width=\columnwidth]{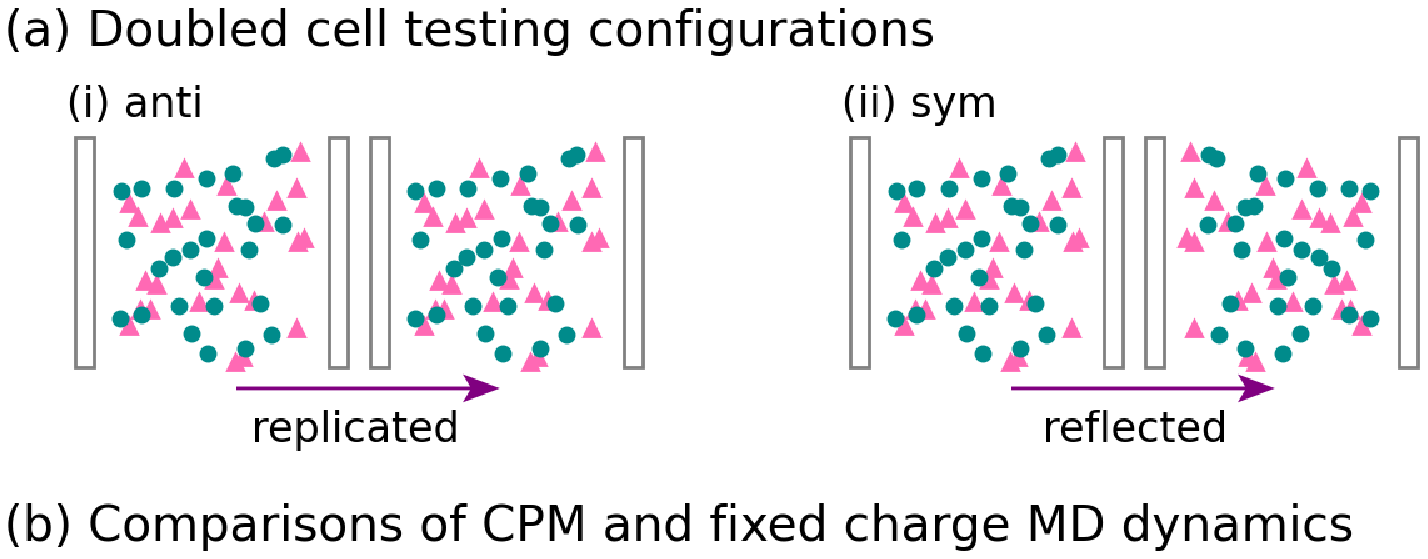}
    \includegraphics[width=\columnwidth]{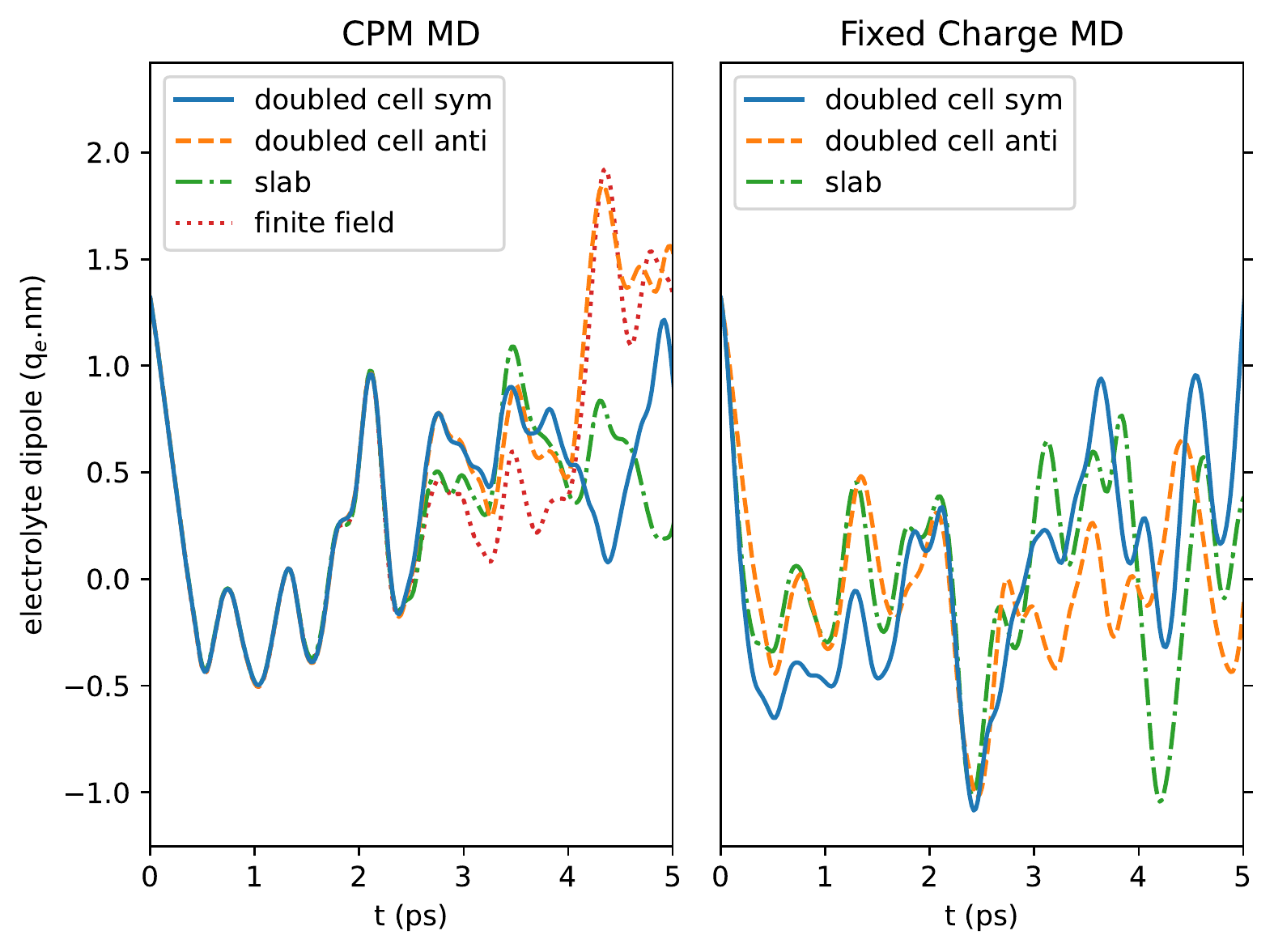}
    \caption{Validating the independence of the two cells in the doubled cell CPM MD with \emph{anti} vs \emph{sym} replication. (a) Depictions of \textit{anti} and \textit{sym} initial configurations. (b) Cell dipole evolution for short trajectories of CPM and fixed charge MD in various unit cell configurations. Single lines are shown for each doubled cell method simulation, as the divergence between individual cell dipoles is not visible on the graph within the 6 ps duration shown (for both CPM and fixed charge MD). \label{fig:trajcheck}}
\end{figure}

To further validate trajectory independence in the doubled cell method, two short trajectories with $\Delta \psi = 0 V$ and with different initial conditions were compared (Figure \ref{fig:trajcheck} (a)). In the \emph{antisymmetric} (or \emph{anti}) initial condition, a single cell configuration (comprising both positions and velocities) was reflected along the $z$-axis, while in the \emph{symmetric} (or \emph{sym}) initial condition the duplicated cell was also $z$-reversed, resulting in a reflected image. Thus, in the \emph{sym} condition, the electrolyte and electrodes initially have the same alignment in both cells, while in the \emph{anti} condition the electrolyte and electrodes initially have opposite alignment, making two maximally different initial conditions.  If the duplicated cell does not influence the behaviour of the original cell, then (to within numerical error) the properties of the system in the first cell will not change if the duplicated cell changes. 

Figure \ref{fig:trajcheck} (b)(i) shows the subsequent evolution of the electrolyte dipole moment ($\sum Q_j Z_j$) in the left cell from these initial configurations, as well as for slab and finite field-based comparison trajectories. Importantly, the divergence between \emph{anti} and \emph{sym} trajectories primarily emerges at about 2.5 ps, showing that until that point the left cell evolves identically despite the right cell being maximally different. The slab, finite field, and doubled cell trajectories start diverging just before then, suggesting that the trajectory divergence is a result of typical floating-point error accumulation. We note that in the doubled cell method, two separate Nose-Hoover thermostats are used, one for each cell; using a single Nose-Hoover thermostat across the electrolytes of both cells couples them together and reduces their independence (data not shown).

We demonstrate that for the fixed charge MD, statistical independence of the two cells is not observed  by repeating simulation of the short trajectories under a fixed charge condition, simply leaving all electrode particles neutral. As seen in Figure \ref{fig:trajcheck} (b)(ii), there is an immediate difference between \emph{sym} and \emph{anti} trajectories, showing that in fixed charge MD the two cells are no longer completely uncoupled. Both the \emph{sym} and \emph{anti} trajectories also quickly diverge from a single cell, slab-corrected fixed charge comparison trajectory.

\subsection{Electrolyte and Charge Densities}
\label{ss:ecd}

\begin{figure}
    \centering
    \includegraphics[width=\columnwidth]{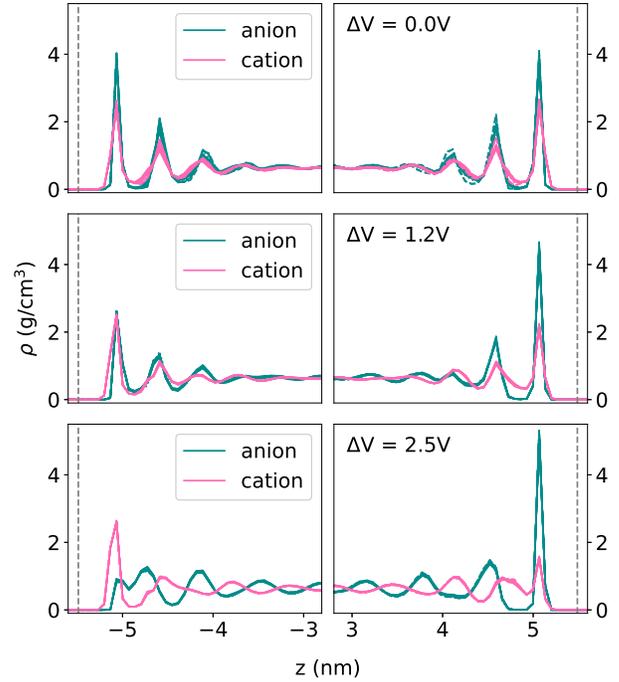}%
    \caption{Trajectory averaged anion and cation densities near the cell electrodes at electrode potential differences of $\Delta V$ = 0.0 V (top), 1.2 V (middle), and 2.5 V (bottom), with the dashed gray lines indicating the positions of the electrodes nearest to the ionic liquid. Densities were drawn using dashed, dotted, and solid lines for slab, finite field, and doubled cell geometries respectively, but the differences cannot be visually distinguished and are not larger than between different runs using identical geometries.\label{fig:cl}}
\end{figure}

Figure \ref{fig:cl} shows the equilibrium density profiles of BMim$^+$ and PF$_6^-$ particles across the cell for the imposed potential differences $\Delta V = $ 0.0, 1.2 and 2.5 V (with graphs for other potential differences supplied in the SI Fig D). Again, little difference is seen between the slab, finite field, and doubled cell geometries. Ionic layers are observed to form at the interfaces with the conducting electrodes, with the layers at 0.0 V attributable purely to size and symmetry of the ions and their interactions with the carbon atoms of the electrode. At higher voltages, distinct phenomena control the electrode ionic layering. The anion, which is represented as a sphere, can be packed into the layer nearest the electrode with increasing density at higher potential differences. Since the cation has an irregular shape, cationic layers cannot pack with the same efficiency, and increased electrode charge causes cationic density to broaden and build up in the second layer instead. In either case, there is a clear change from co-layering at low electrode charges, where anion and cation layers almost coincide, to counter-layering where anions and cations alternate.

\begin{figure}
    \centering
    \includegraphics[width=\columnwidth]{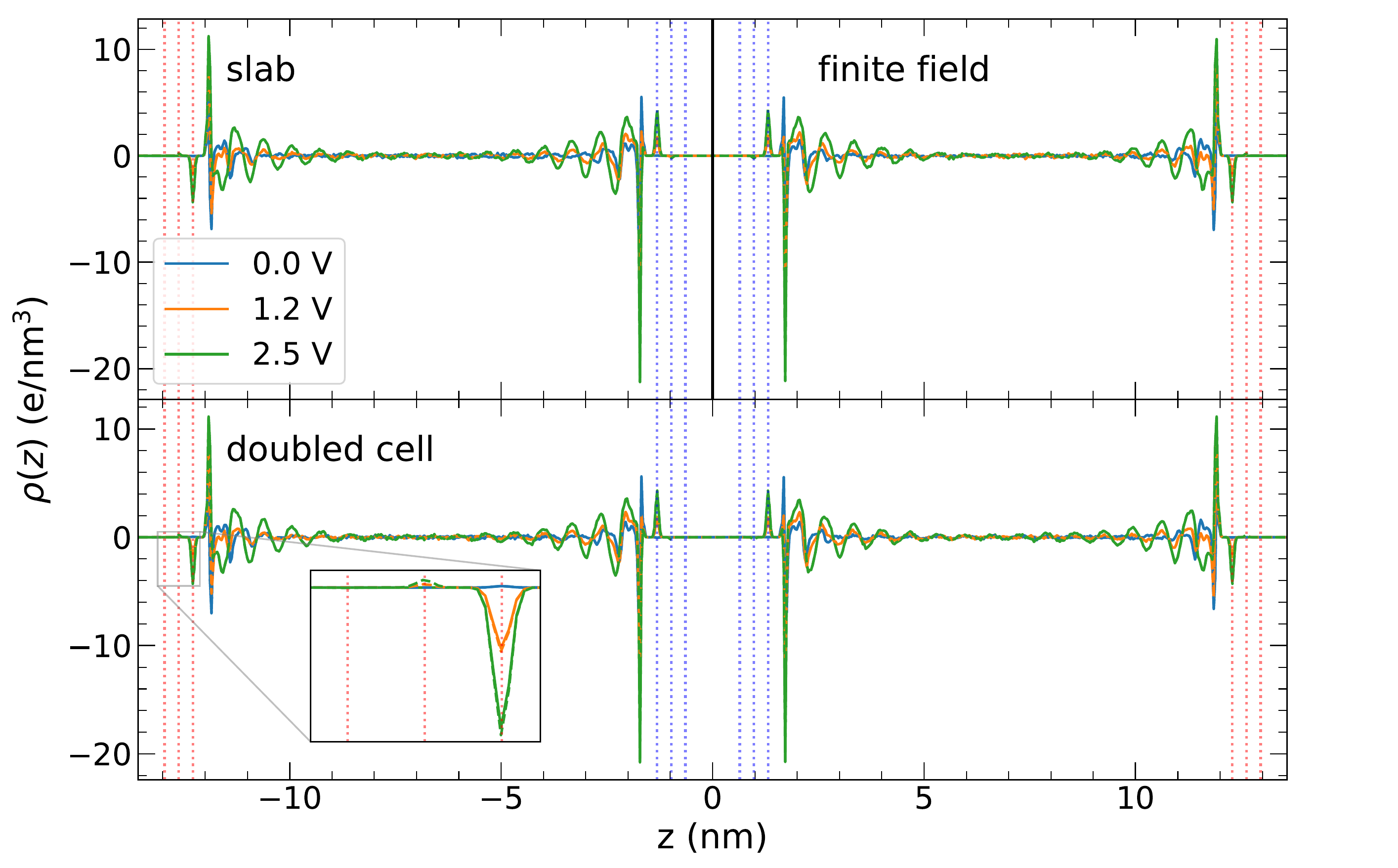}
    \caption{Trajectory averaged charge densities across the simulation cell for slab (top left), finite field (top right, with $z$ reversed), and doubled cell (bottom) geometries, at imposed potential differences $\Delta V =$ 0.0, 1.2, and 2.5 V. Electrode positions are indicated by the dotted lines, with blue (red) colour indicating the layers of the positive (negative) electrodes. (Inset) The electrode region is magnified to better visualize the three-layer predicted charges (dashed lines).\label{fig:qz}}
\end{figure}

Figure \ref{fig:qz} shows the charge density across the cell, accounting for the Gaussian charge densities on the electrode layers. Again, all three methods return very similar charge densities, with larger oscillations near the electrodes as the potential difference increases. Comparing the post-processed three-layer charges to the single-layer charges shows that, when CPM MD is applied to all three electrode layers, more than 90\% of the charge is still induced on the single layers closest to the electrolyte. Leaving the basal layers uncharged is thus expected to have minimal effect on the electrolyte dynamics, but has consequences for obtaining the correct potential profile using Poisson methods, as discussed below.

\subsection{Simulation Cell Potential Profiles and Differential Capacitances}

\begin{figure}
    \centering
    \includegraphics[width=\columnwidth]{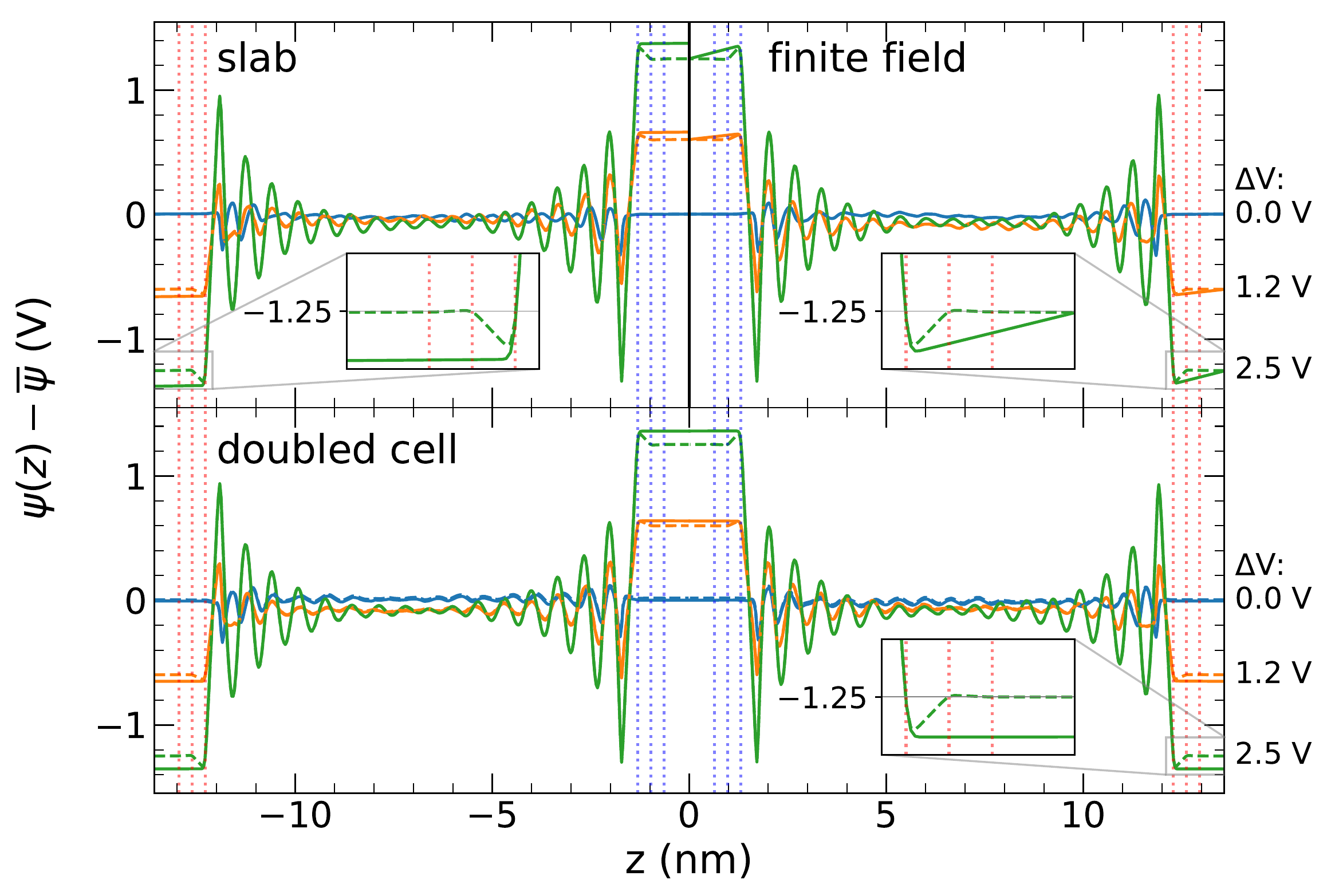}
    \caption{Trajectory averaged electrostatic potential ($\psi(z) - \overline{\psi}$) across the simulation cell for slab (top left), finite field (top right), and doubled cell (bottom) methods, at imposed potential differences $\Delta V =$ 0.0, 1.2, and 2.5 V. As in Figure \ref{fig:qz}, solid lines are single-layer potential profiles and dashed lines are three-layer predicted potential profiles, and electrode positions are indicated by the dotted lines. The solid lines for the slab results are obtained by setting the potential difference between the innermost electrode layers to the simulation-imposed value. (Insets) Potentials in the electrode regions are magnified to better visualize the effect of inner-layer charges on the potential profile.    \label{fig:vz}}
\end{figure}

Figure \ref{fig:vz} shows the potential profiles, $\psi(z) - \overline{\psi}$, across the cell for $\Delta V = $ 0.0, 1.2, and 2.5 V, in slab, finite field, and doubled cell methods. The average values of $\overline{\psi}$ are very small compared to $\psi(z)$ for these systems (e.g. for slab simulations, $|\overline{\psi}|<0.03V$ at all potentials considered).  As in Figure \ref{fig:qz}, solid lines show the single-layer potential profile and dashed lines show the three-layer potential profile. 

The three-layer potential profiles show little difference between the three methods (besides statistical variation between trajectories). For all methods, the potential difference between the inner-most electrodes and the cell boundaries is equal to the imposed electrode potential difference, as with the finite field and doubled cell methods.

The single-layer potential profiles within the electrodes, on the other hand, are visibly affected by the choice of boundary conditions, and differ visibly between slab, finite field, and doubled cell methods. We discuss these results for the particular case of $\Delta \psi = 2.5 $V, but the same phenomena are seen at all other potential differences (shown in Figure \ref{fig:vz} for 0.0 and 1.2 V).

Considering the single-layer charged electrodes first (solid lines in Figure \ref{fig:vz}), the insets show that for all systems $\psi(z) - \overline{\psi}$ is exactly -1.25 V at the electrode in contact with the electrolyte, as set. Between this electrode and the next, the magnitude of the potential initially increases in all cases, due to the Gaussian distribution of the charges.  In the slab and double cell system, it then reaches a constant value which is maintained until the edget of the 2D simulation cell is reached (slab) or the charge of the other surface electrode becomes significant (doubled cell).  In the finite field method, the field due to the added ramp potential that has been applied across the whole simulation cell rather than between the electrodes is evident, and results in a linear drop in the magnitude back to the value of -1.25 V at the simulation cell boundary.  For all systems with use of the three-layer charged electrodes, $\psi(z) - \overline{\psi}$ at each of the three electrodes is fixed to -1.25 V, so after the initial increase in magnitude of the electrostatic potential due to the Gaussian distribution of charge, there is a drop back to -1.25 V at the next electrode. Since the charge on this electrode is small, there is little evidence of the Gaussian distribution of charges. The potential within the electrode is artificially increased due to the absence of the counter-charges from the inner layers.

\begin{figure}
    \centering
    \includegraphics[width=\columnwidth]{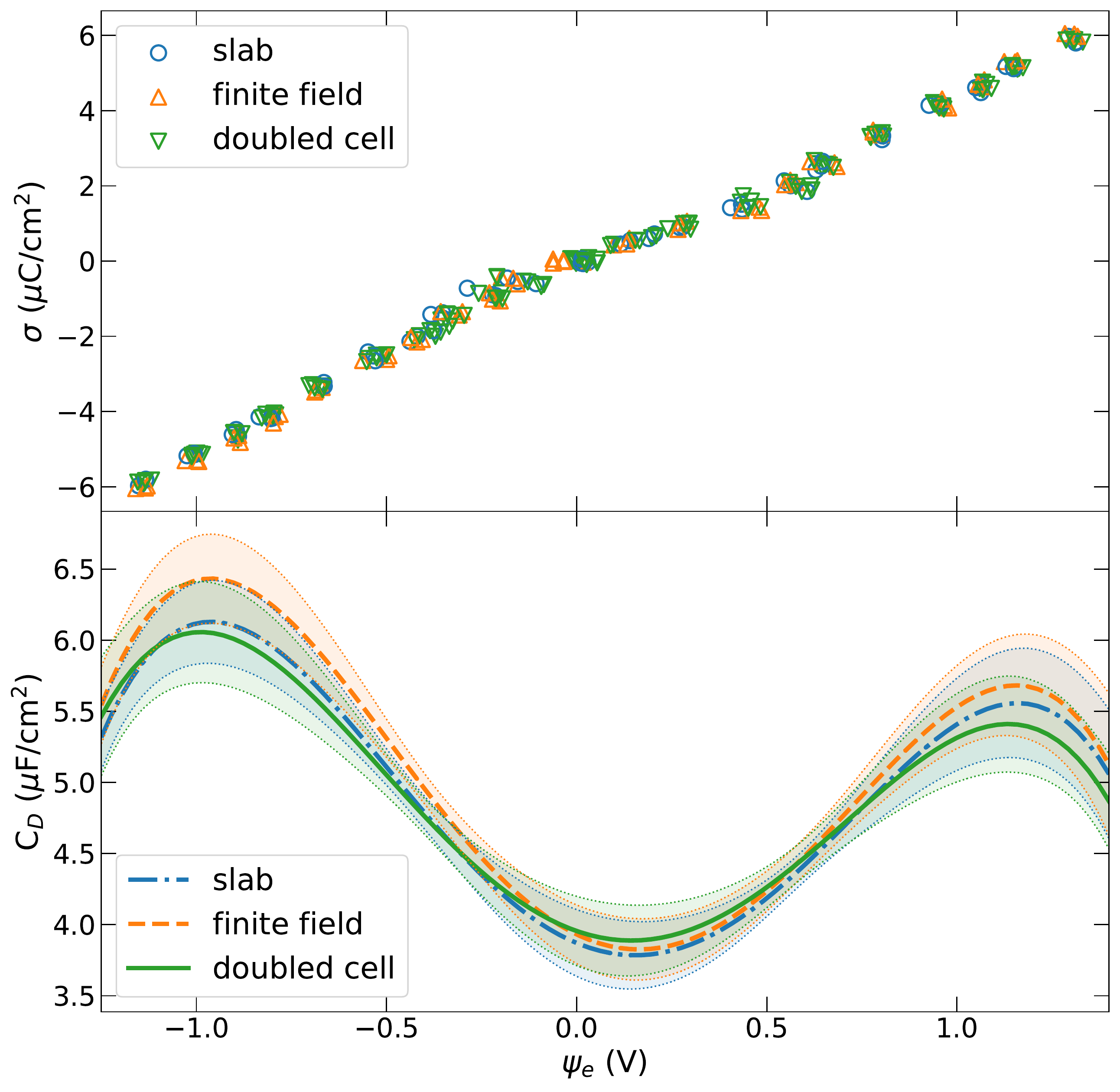}
    \caption{Electrode charges (top) and differential capacitance (bottom) as a function of the electrode potential $\psi_e$, calculated from three-layer charges.
    \label{fig:dc}}
\end{figure}

Given the electrostatic potential profiles, the electric surface potential (the difference in the electrostatic potential at each electrode and in the bulk,) $\psi_e$ can be determined, and the dependence of the electrode charges on $\psi_e$  are shown in Figure \ref{fig:dc}. Within statistical error, the same average charge and electrode potentials are evolved for each simulation at the same potential difference, regardless of the method used. Therefore the calculated differential capacitance, $C_D = d\sigma/d\psi_e$,  is also the same between slab, finite field and doubled cell methods (within statistical uncertainties), as Figure \ref{fig:dc} shows, and the  double-humped curve characteristic of screening and overcrowding in complex ionic electrolytes is obtained \cite{Kornyshev2007Double-layerChange,Bazant2009TowardsSolutions}. The method used to determine the uncertainties in $C_D$ is discussed in the Supplementary Information.

\subsection{Computational Efficiency}

\begin{figure}
    \centering
    \includegraphics[width=\columnwidth]{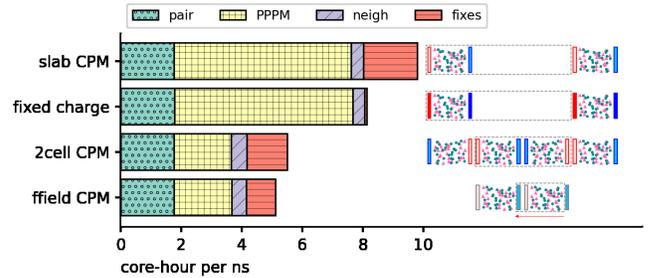}
    \caption{CPU-hours per nanosecond for simulating the ionic liquid-electrode system in slab, fixed charge, finite field, and doubled cell geometries, on four Intel Xeon Haswell 2.6 GHz processors. The time taken is itemized by pair interactions, PPPM long-range electrostatic calculations, neighbor-list building, and LAMMPS ``fixes'' -- mainly the additional computation required for constant potential routines. Right-hand diagrams show each geometry as depicted in Figure \ref{fig:configandsnapshot}.  \label{fig:comp}}
\end{figure}

Although the three methods compared so far give similar results, their computational costs required differ greatly. Figure \ref{fig:comp} shows the computational expense for simulations using each method running on four Intel Xeon Haswell 2.6 GHz processors. (For the doubled cell geometry, each nanosecond of simulation time was counted as providing two nanoseconds of simulation data, since sub-cells have independent dynamics.) The supercapacitor system was also simulated in slab geometry with fixed (single-layer) charges of $+/- 9 \times 10^{-5} \,e$ per atom, to allow the computational costs of CPM and FCM MD to be compared.

Comparison of constant potential and fixed charge methods for the slab geometry shows that CPM charge updating incurs a computational overhead of about 20\% for this system. A similar overhead is incurred in doubled cell and finite field geometries -- however, the significantly reduced unit cell size, and subsequent reduction in long-range electrostatic calculations, more than offsets the overhead. In either periodic geometry the long-range electrostatic calculations are about 60\% cheaper than in the slab geometry. As such, on aggregate, either fully-periodic method for CPM MD results in simulations that are 30--35\% \textit{cheaper} than FCM MD slab simulations. The strong scaling (speedup when using more processors on a problem of the same size) is similar for all four methods, with  the finite field and doubled cell methods showing slightly better scaling.

In closing, it is interesting to note that the use of doubled-cell configuration has been used in fixed charge MD simulations for some time, with the purpose of more realistically applying a field across a slab-like system\cite{Raiteri2020MolecularInterface}. One main purpose of our paper is to demonstrate the application of the doubled-cell configuration to CPM MD simulations, which are important for modelling realistic electrode-electrolyte interactions, but our results suggest that in a doubled-cell, fixed charge simulation the two sub-cells are not fully decoupled (while in a CPM MD simulation they are). Fixed charge simulations will generally be more straightforward and quicker than an equivalent CPM MD simulation, and if the quantities of interest do not depend intimately on the details of electrode charges (such as polarization in the bulk electrolyte, far from electrodes) then the added complexity of a CPM MD simulation may not be worthwhile. Nonetheless, our results show that the novel application of a doubled-cell configuration to CPM MD can result in substantial computational savings and true decoupling of the trajectories of both sub-cells.

\section{Conclusions}

We have shown that fully periodic geometries are useful for accelerating CPM MD simulations of electrodes and their interactions with ionic liquid electrolytes. The resulting efficiency gains in long-range electrostatic calculations can more than offset the cost of the CPM charge update procedure, resulting in CPM MD simulations that are computationally cheaper than their fixed charge equivalents in slab geometries. 

We have demonstrated these capabilities of CPM MD in a computational ionic liquid supercapacitor with flat electrodes. The charging behavior over time, ionic and charge densities across the cell, and resulting estimates of differential capacitance are statistically identical between the slab geometry and the two periodic geometries tested, namely finite field and doubled cell. In doubled cell geometry, the electrode charges evolved in each sub-cell are statistically uncorrelated, and the dynamics of each sub-cell are found to be independent of the electrolyte configuration of the other sub-cell.

When deriving the simulation cell potential profile in order to calculate electrode potentials and differential capacitances, the use of fully periodic geometries also entails simplified boundary conditions which make trajectory post-processing easier. In our study, we found the use of single-layer charges had no effect on the bulk potential, despite multiple charged sheets being physically necessary to screen charge within the electrode. Thus, the significant advantages of fully periodic CPM MD argue for its wider adoption in simulating electrode-electrolyte interactions.

\begin{acknowledgements}

The authors thank the Australian Research Council for its support for this project through the Discovery program (DP180104031 and FL190100080). We would like to thank Dr Emily Kahl for her invaluable support in developing and debugging the source code used in this project. We acknowledge access to computational resources at the NCI National Facility through the National Computational Merit Allocation Scheme supported by the Australian Government, and this work was also supported by resources provided by the Pawsey Supercomputing Centre with funding from the Australian Government and the government of Western Australia. We also acknowledge support from the Queensland Cyber Infrastructure Foundation (QCIF) and the University of Queensland Research Computing Centre (RCC).

\end{acknowledgements}

\section{Appendix}
\label{ss:simdetails}

\subsection{Force Field and Overall Simulation Details}

The electrolyte modelled was an ionic liquid, [BMim$^+$][PF$_6^-$], using the coarse-grained force field of Roy and Maroncelli \cite{Roy2010AnModel}, with three-site cations (kept rigid using SHAKE) and one-site anions. Each electrode was modelled as three graphene sheets with the usual A-B-A staggering, bond-bond (1.42 \AA) and interlayer (3.35 \AA) distances, and the Lennard-Jones parameters of Cole and Klein \cite{Cole1983TheGraphite}; this combination of CG IL and graphene force fields has frequently been used in prior research \cite{Merlet2011ImidazoliumSimulations}. Non-Coulombic interactions were modelled with the Lennard-Jones form with a cutoff of 16 \AA, whereas Coulombic interactions were modelled using particle-particle particle-mesh (PPPM) summation \cite{Hockney1988ComputerParticles} to a relative accuracy of $10^{-8}$.


All MD simulations were integrated using a velocity-Verlet algorithm with a time step of 2 fs. A Nose-Hoover thermostat \cite{Nose1984AEnsemble,Nose1984AMethods,Hoover1985CanonicalDistributions} with a time constant of 100 fs was applied to the electrolyte particles to maintain a temperature of 400K, and cations were kept rigid using the SHAKE algorithm \cite{Ryckaert1977NumericalN-alkanes}. The LAMMPS package \cite{LAMMPS} was used to run simulations, with modifications to implement the CPM charge update algorithms. The extra code was based on the prior package LAMMPS-CONP \cite{Wang2014EvaluationCapacitors} with further optimizations, and is freely available on GitHub.

\subsection{Equilibration, Production, and Post-Analysis}
\label{ss:epp}

Bulk simulations of 1440 ion pairs were first conducted for 4 ns using an NPT barostat \cite{Shinoda2004RapidStress} at 1 bar with a time constant of 4 ps, and the bulk density of the CG IL model was determined to be 1.267 g cm$^{-3}$. A 320 ion pair lattice was then initialized and equilibrated under $x$- and $y$- periodic boundary conditions, with cell sides 32.2 \AA{} and 34.4 \AA{} respectively, while wall potentials with the electrode Lennard-Jones parameters were applied in the $z$-direction until bulk density was replicated in the middle half of the configuration over 1 ns. This slab configuration was then combined with a pair of electrodes whose distance was scanned to maintain bulk density, resulting in a final distance between proximal electrode planes of 109.75 \AA{} and a unit cell z length of 136 \AA. 

From this initial state, CPM MD simulations were run for 30 ns ($1.5 \times 10^7$ time steps) in either slab, single cell (finite field), or doubled cell geometries; doubled cell initial states were formed by replicating the one-cell initial state in the $z$-direction and then flipping positions of electrolytes in the second sub-cell. For each geometry, 11 potential differences were used (0.0, 0.3, 0.5, 0.8, 1.0, 1.2, 1.5, 1.8, 2.0, 2.2, and 2.5 V). Electrode charges were updated every 5 steps (10 fs), which is acceptable since other studies report accurate results even with less frequent charge updates \cite{Tu2020InnerDynamics}, and the electrode charges were modelled as Gaussian distributions   (see (\ref{eqn:gauss_definition})) with $\eta = 1.979/$\AA, as used in other studies \cite{Wang2014EvaluationCapacitors}. After each set of simulations had been completed, the final state of the 0.0 V simulation was used as the starting state for a new set of simulations; this was repeated twice for a total of three independent simulation sets. At an accuracy of $10^{-8}$, the PPPM meshes used for slab, finite-field and doubled cell geometries contained $30 (x)\times30 (y)\times225(z) $, $30\times30\times90$ and $30\times32\times180$ grid points respectively.

For each run, position configurations were written to disk every 20 ps ($10^4$ time steps). During each run, only the electrode layers nearest to the ionic liquid were charged with the CPM update procedure, while the further two layers of each electrode were left neutral and only contributed Lennard-Jones interactions. To study the accuracy of this approximation, the snapshots of each trajectory were re-run, and the charges that would have evolved had all three electrode layers been charged were recorded for analysis. As described in Section \ref{ss:ecd}, the resulting charges on the further two layers are very small and unlikely to significantly affect the observed dynamics, but they can affect the calculation of the cell potential profile.

The steady state particle and charge densities were subsequently obtained over a $z$-grid spacing of 0.34 \AA, corresponding to 400 grid points per unit cell for the slab and finite field geometries and 800 grid points per unit cell for the doubled cell geometries, and the electrostatic potential profile $\psi(z)$ was obtained using finite differences as described in the text. The average value of $\psi(z)$ across the middle 100 grid points of each unit cell (sub-cell, for doubled-cell calculations) was then defined as the bulk potential and set to $0 V$ for calculating the anode and cathode potentials.

Each set of runs thus contributed 22 data points (two for each potential difference) to the plot of electrode charge against potential in Figure \ref{fig:dc}. Charge-against-potential data sets were subsequently used to estimate the differential capacitance, $C_D = d\sigma/d\psi_e$, by fitting the data set to a fifth-order spline between --1.2 and 1.3 V. The gradient at each end-point was constrained to be equal to the linear least squares gradient of the five furthest points, to prevent oscillatory overfitting at the end points. The uncertainty in $C_D$ was estimated by bootstrapping: each of the 22 data points in the charge-potential curve could take one of three possible values (six, for the doubled-cell results) given the three independent sets of runs, and thus independent $C_D$ curves could be calculated based on which run was chosen at which point. 200 such independent $C_D$ curves were calculated and the 95\% confidence interval, shown in Figure \ref{fig:dc}, was chosen as 1.96 $\times$ the standard deviation at each potential.

\subsection{Initial Charge Trajectories, Charge Density Profiles and Potential Profiles for All Potential Differences}

The following figures show initial charge trajectories, charge density profiles, and potential profiles for all potential differences. While the data shown here is derived from the slab trajectories, similar results are seen when visualising the same quantities from the finite field and doubled cell trajectories.

\begin{figure}
    \centering
    \includegraphics[width=0.8\columnwidth]{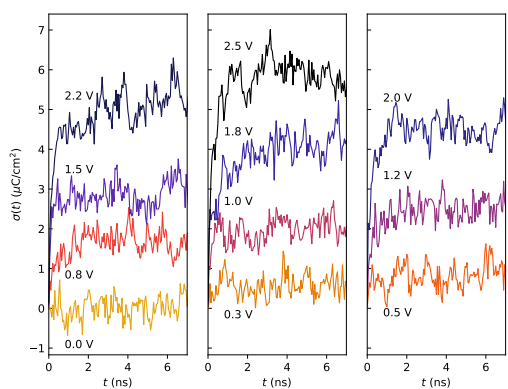}
    \caption{Typical trajectories of instantaneous surface charge density against simulation time for all potential differences studied under CPM MD, spread over three panels for clarity. The set shown was collected using slab simulations; as discussed in the main text, finite field and doubled cell simulations yield similar results. Each trajectory is labeled with the potential difference used, and the lines are also color-coded using the same color scheme as subsequent graphs. The graphs are spread over three different panels for better visibility \label{fig:qt_allv}}
\end{figure}

\begin{figure}
    \centering
    \includegraphics[width=\columnwidth]{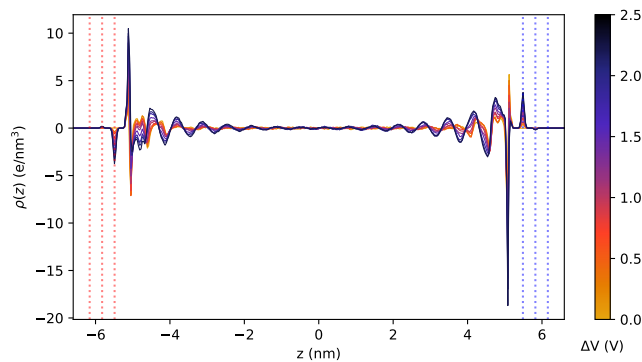}
    \caption{Single cell charge profiles for all potential differences studied, with the positions of negative (positive) electrodes denoted by red (blue) dotted lines as in the main text figures, and different line colors showing the simulation imposed potential difference. The three-layer charge is depicted here, but the charges induced on the two further layers are visibly negligible.\label{fig:qallv}}
\end{figure}

\begin{figure}
    \centering
    \includegraphics[width=\columnwidth]{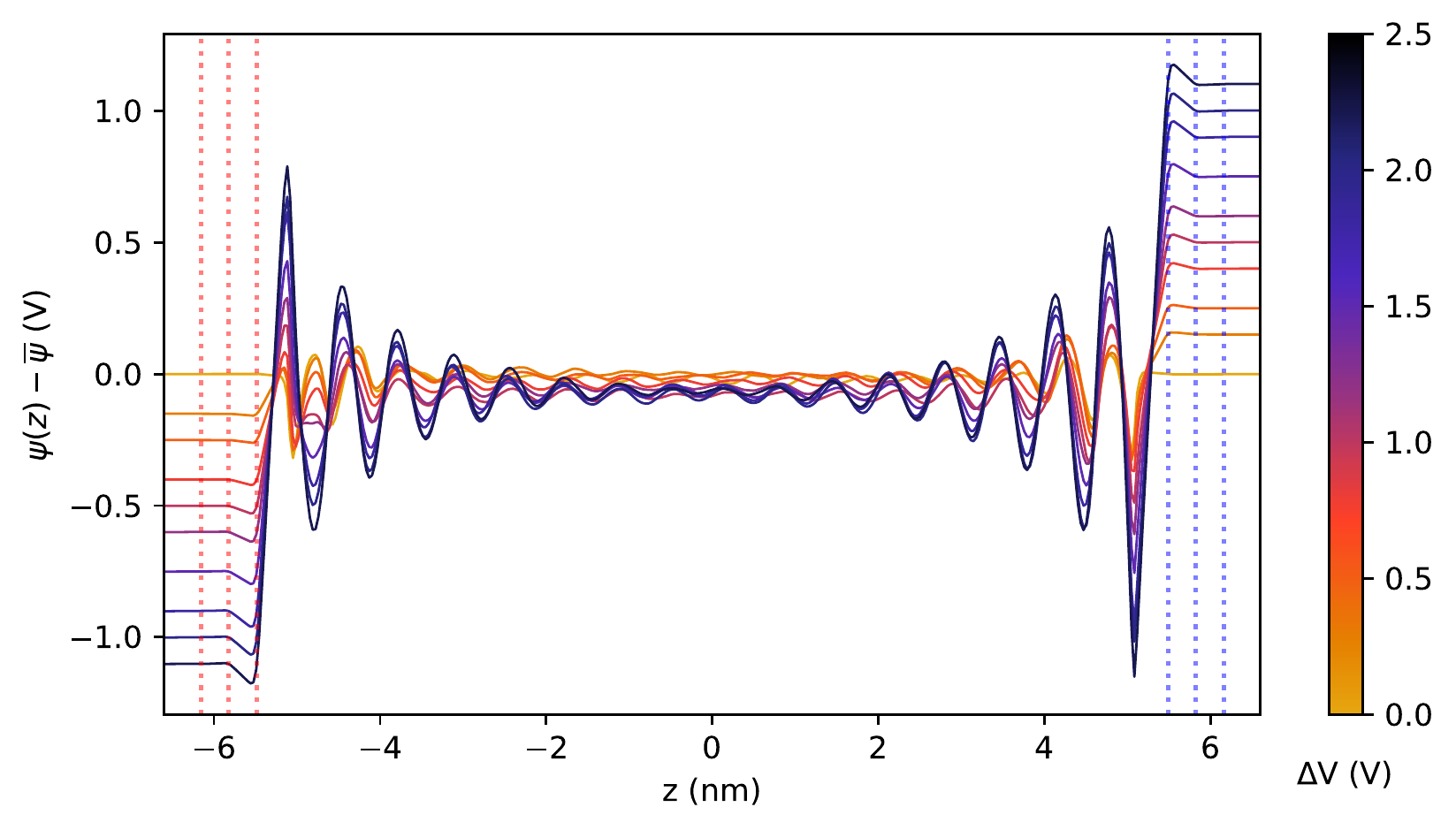}
    \caption{Single cell potential profiles for all potential differences studied, with the positions of negative (positive) electrodes denoted by red (blue) dotted lines as in the main text figures, and different line colors showing the simulation imposed potential difference. The three-layer potentials are depicted here and a constant shift is applied so that the potentials at the cell edges are $\pm \Delta V/2$. The downshifting of the bulk potential at higher potential differences is noticeable and corresponds to the asymmetric electrode differential capacitance documented in the text.\label{fig:phiallv}}
\end{figure}

\newpage
\section{References}
\nocite{*}

\bibliography{onlyrefs}

\begin{thebibliography}{50}%
\makeatletter
\providecommand \@ifxundefined [1]{%
 \@ifx{#1\undefined}
}%
\providecommand \@ifnum [1]{%
 \ifnum #1\expandafter \@firstoftwo
 \else \expandafter \@secondoftwo
 \fi
}%
\providecommand \@ifx [1]{%
 \ifx #1\expandafter \@firstoftwo
 \else \expandafter \@secondoftwo
 \fi
}%
\providecommand \natexlab [1]{#1}%
\providecommand \enquote  [1]{``#1''}%
\providecommand \bibnamefont  [1]{#1}%
\providecommand \bibfnamefont [1]{#1}%
\providecommand \citenamefont [1]{#1}%
\providecommand \href@noop [0]{\@secondoftwo}%
\providecommand \href [0]{\begingroup \@sanitize@url \@href}%
\providecommand \@href[1]{\@@startlink{#1}\@@href}%
\providecommand \@@href[1]{\endgroup#1\@@endlink}%
\providecommand \@sanitize@url [0]{\catcode `\\12\catcode `\$12\catcode
  `\&12\catcode `\#12\catcode `\^12\catcode `\_12\catcode `\%12\relax}%
\providecommand \@@startlink[1]{}%
\providecommand \@@endlink[0]{}%
\providecommand \url  [0]{\begingroup\@sanitize@url \@url }%
\providecommand \@url [1]{\endgroup\@href {#1}{\urlprefix }}%
\providecommand \urlprefix  [0]{URL }%
\providecommand \Eprint [0]{\href }%
\providecommand \doibase [0]{http://dx.doi.org/}%
\providecommand \selectlanguage [0]{\@gobble}%
\providecommand \bibinfo  [0]{\@secondoftwo}%
\providecommand \bibfield  [0]{\@secondoftwo}%
\providecommand \translation [1]{[#1]}%
\providecommand \BibitemOpen [0]{}%
\providecommand \bibitemStop [0]{}%
\providecommand \bibitemNoStop [0]{.\EOS\space}%
\providecommand \EOS [0]{\spacefactor3000\relax}%
\providecommand \BibitemShut  [1]{\csname bibitem#1\endcsname}%
\let\auto@bib@innerbib\@empty
\bibitem [{\citenamefont {Liu}\ \emph {et~al.}(2010)\citenamefont {Liu},
  \citenamefont {Liu},\ and\ \citenamefont
  {Li}}]{Liu2010IonicElectrochemistry}%
  \BibitemOpen
  \bibfield  {author} {\bibinfo {author} {\bibfnamefont {H.}~\bibnamefont
  {Liu}}, \bibinfo {author} {\bibfnamefont {Y.}~\bibnamefont {Liu}}, \ and\
  \bibinfo {author} {\bibfnamefont {J.}~\bibnamefont {Li}},\ }\href {\doibase
  10.1039/b921469k} {\bibfield  {journal} {\bibinfo  {journal} {Physical
  Chemistry Chemical Physics}\ }\textbf {\bibinfo {volume} {12}},\ \bibinfo
  {pages} {1685} (\bibinfo {year} {2010})}\BibitemShut {NoStop}%
\bibitem [{\citenamefont {Lian}\ \emph {et~al.}(2019)\citenamefont {Lian},
  \citenamefont {Liu}, \citenamefont {Li},\ and\ \citenamefont
  {Wu}}]{Lian2019HuntingWindows}%
  \BibitemOpen
  \bibfield  {author} {\bibinfo {author} {\bibfnamefont {C.}~\bibnamefont
  {Lian}}, \bibinfo {author} {\bibfnamefont {H.}~\bibnamefont {Liu}}, \bibinfo
  {author} {\bibfnamefont {C.}~\bibnamefont {Li}}, \ and\ \bibinfo {author}
  {\bibfnamefont {J.}~\bibnamefont {Wu}},\ }\href {\doibase 10.1002/aic.16467}
  {\bibfield  {journal} {\bibinfo  {journal} {AIChE Journal}\ }\textbf
  {\bibinfo {volume} {65}},\ \bibinfo {pages} {804} (\bibinfo {year}
  {2019})}\BibitemShut {NoStop}%
\bibitem [{\citenamefont {Hayes}\ \emph {et~al.}(2015)\citenamefont {Hayes},
  \citenamefont {Warr},\ and\ \citenamefont
  {Atkin}}]{Hayes2015StructureLiquids}%
  \BibitemOpen
  \bibfield  {author} {\bibinfo {author} {\bibfnamefont {R.}~\bibnamefont
  {Hayes}}, \bibinfo {author} {\bibfnamefont {G.~G.}\ \bibnamefont {Warr}}, \
  and\ \bibinfo {author} {\bibfnamefont {R.}~\bibnamefont {Atkin}},\ }\href
  {\doibase 10.1021/cr500411q} {\bibfield  {journal} {\bibinfo  {journal}
  {Chemical Reviews}\ }\textbf {\bibinfo {volume} {115}},\ \bibinfo {pages}
  {6357} (\bibinfo {year} {2015})}\BibitemShut {NoStop}%
\bibitem [{\citenamefont {Kornyshev}(2007)}]{Kornyshev2007Double-layerChange}%
  \BibitemOpen
  \bibfield  {author} {\bibinfo {author} {\bibfnamefont {A.~A.}\ \bibnamefont
  {Kornyshev}},\ }\href {\doibase 10.1021/jp067857o} {\bibfield  {journal}
  {\bibinfo  {journal} {Journal of Physical Chemistry B}\ }\textbf {\bibinfo
  {volume} {111}},\ \bibinfo {pages} {5545} (\bibinfo {year}
  {2007})}\BibitemShut {NoStop}%
\bibitem [{\citenamefont {Bedrov}\ \emph {et~al.}(2019)\citenamefont {Bedrov},
  \citenamefont {Piquemal}, \citenamefont {Borodin}, \citenamefont {MacKerell},
  \citenamefont {Roux},\ and\ \citenamefont
  {Schr{\"{o}}der}}]{Bedrov2019MolecularFields}%
  \BibitemOpen
  \bibfield  {author} {\bibinfo {author} {\bibfnamefont {D.}~\bibnamefont
  {Bedrov}}, \bibinfo {author} {\bibfnamefont {J.~P.}\ \bibnamefont
  {Piquemal}}, \bibinfo {author} {\bibfnamefont {O.}~\bibnamefont {Borodin}},
  \bibinfo {author} {\bibfnamefont {A.~D.}\ \bibnamefont {MacKerell}}, \bibinfo
  {author} {\bibfnamefont {B.}~\bibnamefont {Roux}}, \ and\ \bibinfo {author}
  {\bibfnamefont {C.}~\bibnamefont {Schr{\"{o}}der}},\ }\href {\doibase
  10.1021/acs.chemrev.8b00763} {\bibfield  {journal} {\bibinfo  {journal}
  {Chemical Reviews}\ }\textbf {\bibinfo {volume} {119}},\ \bibinfo {pages}
  {7940} (\bibinfo {year} {2019})}\BibitemShut {NoStop}%
\bibitem [{\citenamefont {Doherty}\ \emph {et~al.}(2017)\citenamefont
  {Doherty}, \citenamefont {Zhong}, \citenamefont {Gathiaka}, \citenamefont
  {Li},\ and\ \citenamefont {Acevedo}}]{Doherty2017RevisitingSimulations}%
  \BibitemOpen
  \bibfield  {author} {\bibinfo {author} {\bibfnamefont {B.}~\bibnamefont
  {Doherty}}, \bibinfo {author} {\bibfnamefont {X.}~\bibnamefont {Zhong}},
  \bibinfo {author} {\bibfnamefont {S.}~\bibnamefont {Gathiaka}}, \bibinfo
  {author} {\bibfnamefont {B.}~\bibnamefont {Li}}, \ and\ \bibinfo {author}
  {\bibfnamefont {O.}~\bibnamefont {Acevedo}},\ }\href {\doibase
  10.1021/acs.jctc.7b00520} {\bibfield  {journal} {\bibinfo  {journal} {Journal
  of Chemical Theory and Computation}\ }\textbf {\bibinfo {volume} {13}},\
  \bibinfo {pages} {6131} (\bibinfo {year} {2017})}\BibitemShut {NoStop}%
\bibitem [{\citenamefont {Roy}\ and\ \citenamefont
  {Maroncelli}(2010)}]{Roy2010AnModel}%
  \BibitemOpen
  \bibfield  {author} {\bibinfo {author} {\bibfnamefont {D.}~\bibnamefont
  {Roy}}\ and\ \bibinfo {author} {\bibfnamefont {M.}~\bibnamefont
  {Maroncelli}},\ }\href {\doibase 10.1021/jp108179n} {\bibfield  {journal}
  {\bibinfo  {journal} {Journal of Physical Chemistry B}\ }\textbf {\bibinfo
  {volume} {114}},\ \bibinfo {pages} {12629} (\bibinfo {year}
  {2010})}\BibitemShut {NoStop}%
\bibitem [{\citenamefont {Fajardo}\ \emph {et~al.}(2020)\citenamefont
  {Fajardo}, \citenamefont {Di~Lecce},\ and\ \citenamefont
  {Bresme}}]{Fajardo2020MolecularForce-field}%
  \BibitemOpen
  \bibfield  {author} {\bibinfo {author} {\bibfnamefont {O.~Y.}\ \bibnamefont
  {Fajardo}}, \bibinfo {author} {\bibfnamefont {S.}~\bibnamefont {Di~Lecce}}, \
  and\ \bibinfo {author} {\bibfnamefont {F.}~\bibnamefont {Bresme}},\ }\href
  {\doibase 10.1039/c9cp05932f} {\bibfield  {journal} {\bibinfo  {journal}
  {Physical Chemistry Chemical Physics}\ }\textbf {\bibinfo {volume} {22}},\
  \bibinfo {pages} {1682} (\bibinfo {year} {2020})}\BibitemShut {NoStop}%
\bibitem [{\citenamefont {Siepmann}\ and\ \citenamefont
  {Sprik}(1995)}]{Siepmann1995InfluenceSystems}%
  \BibitemOpen
  \bibfield  {author} {\bibinfo {author} {\bibfnamefont {J.~I.}\ \bibnamefont
  {Siepmann}}\ and\ \bibinfo {author} {\bibfnamefont {M.}~\bibnamefont
  {Sprik}},\ }\href {\doibase 10.1063/1.469429} {\bibfield  {journal} {\bibinfo
   {journal} {The Journal of Chemical Physics}\ }\textbf {\bibinfo {volume}
  {102}},\ \bibinfo {pages} {511} (\bibinfo {year} {1995})}\BibitemShut
  {NoStop}%
\bibitem [{\citenamefont {Reed}\ \emph {et~al.}(2007)\citenamefont {Reed},
  \citenamefont {Lanning},\ and\ \citenamefont
  {Madden}}]{Reed2007ElectrochemicalElectrode}%
  \BibitemOpen
  \bibfield  {author} {\bibinfo {author} {\bibfnamefont {S.~K.}\ \bibnamefont
  {Reed}}, \bibinfo {author} {\bibfnamefont {O.~J.}\ \bibnamefont {Lanning}}, \
  and\ \bibinfo {author} {\bibfnamefont {P.~A.}\ \bibnamefont {Madden}},\
  }\href {\doibase 10.1063/1.2464084} {\bibfield  {journal} {\bibinfo
  {journal} {Journal of Chemical Physics}\ }\textbf {\bibinfo {volume} {126}},\
  \bibinfo {pages} {084704} (\bibinfo {year} {2007})}\BibitemShut {NoStop}%
\bibitem [{\citenamefont {Gingrich}\ and\ \citenamefont
  {Wilson}(2010)}]{Gingrich2010OnSurfaces}%
  \BibitemOpen
  \bibfield  {author} {\bibinfo {author} {\bibfnamefont {T.~R.}\ \bibnamefont
  {Gingrich}}\ and\ \bibinfo {author} {\bibfnamefont {M.}~\bibnamefont
  {Wilson}},\ }\href {\doibase 10.1016/j.cplett.2010.10.010} {\bibfield
  {journal} {\bibinfo  {journal} {Chemical Physics Letters}\ }\textbf {\bibinfo
  {volume} {500}},\ \bibinfo {pages} {178} (\bibinfo {year}
  {2010})}\BibitemShut {NoStop}%
\bibitem [{\citenamefont {Tazi}\ \emph {et~al.}(2010)\citenamefont {Tazi},
  \citenamefont {Salanne}, \citenamefont {Simon}, \citenamefont {Turq},
  \citenamefont {Pounds},\ and\ \citenamefont
  {Madden}}]{Tazi2010Potential-inducedInterface}%
  \BibitemOpen
  \bibfield  {author} {\bibinfo {author} {\bibfnamefont {S.}~\bibnamefont
  {Tazi}}, \bibinfo {author} {\bibfnamefont {M.}~\bibnamefont {Salanne}},
  \bibinfo {author} {\bibfnamefont {C.}~\bibnamefont {Simon}}, \bibinfo
  {author} {\bibfnamefont {P.}~\bibnamefont {Turq}}, \bibinfo {author}
  {\bibfnamefont {M.}~\bibnamefont {Pounds}}, \ and\ \bibinfo {author}
  {\bibfnamefont {P.~A.}\ \bibnamefont {Madden}},\ }\href {\doibase
  10.1021/jp1030448} {\bibfield  {journal} {\bibinfo  {journal} {Journal of
  Physical Chemistry B}\ }\textbf {\bibinfo {volume} {114}},\ \bibinfo {pages}
  {8453} (\bibinfo {year} {2010})}\BibitemShut {NoStop}%
\bibitem [{\citenamefont {Wang}\ \emph {et~al.}(2014)\citenamefont {Wang},
  \citenamefont {Yang}, \citenamefont {Olmsted}, \citenamefont {Asta},\ and\
  \citenamefont {Laird}}]{Wang2014EvaluationCapacitors}%
  \BibitemOpen
  \bibfield  {author} {\bibinfo {author} {\bibfnamefont {Z.}~\bibnamefont
  {Wang}}, \bibinfo {author} {\bibfnamefont {Y.}~\bibnamefont {Yang}}, \bibinfo
  {author} {\bibfnamefont {D.~L.}\ \bibnamefont {Olmsted}}, \bibinfo {author}
  {\bibfnamefont {M.}~\bibnamefont {Asta}}, \ and\ \bibinfo {author}
  {\bibfnamefont {B.~B.}\ \bibnamefont {Laird}},\ }\href {\doibase
  10.1063/1.4899176} {\bibfield  {journal} {\bibinfo  {journal} {Journal of
  Chemical Physics}\ }\textbf {\bibinfo {volume} {141}},\ \bibinfo {pages}
  {184102} (\bibinfo {year} {2014})}\BibitemShut {NoStop}%
\bibitem [{\citenamefont {Haskins}\ and\ \citenamefont
  {Lawson}(2016)}]{Haskins2016EvaluationLayers}%
  \BibitemOpen
  \bibfield  {author} {\bibinfo {author} {\bibfnamefont {J.~B.}\ \bibnamefont
  {Haskins}}\ and\ \bibinfo {author} {\bibfnamefont {J.~W.}\ \bibnamefont
  {Lawson}},\ }\href {\doibase 10.1063/1.4948938} {\bibfield  {journal}
  {\bibinfo  {journal} {Journal of Chemical Physics}\ }\textbf {\bibinfo
  {volume} {144}},\ \bibinfo {pages} {184707} (\bibinfo {year}
  {2016})}\BibitemShut {NoStop}%
\bibitem [{\citenamefont {Xing}\ \emph {et~al.}(2013)\citenamefont {Xing},
  \citenamefont {Vatamanu}, \citenamefont {Borodin},\ and\ \citenamefont
  {Bedrov}}]{Xing2013OnPores}%
  \BibitemOpen
  \bibfield  {author} {\bibinfo {author} {\bibfnamefont {L.}~\bibnamefont
  {Xing}}, \bibinfo {author} {\bibfnamefont {J.}~\bibnamefont {Vatamanu}},
  \bibinfo {author} {\bibfnamefont {O.}~\bibnamefont {Borodin}}, \ and\
  \bibinfo {author} {\bibfnamefont {D.}~\bibnamefont {Bedrov}},\ }\href
  {\doibase 10.1021/jz301782f} {\bibfield  {journal} {\bibinfo  {journal}
  {Journal of Physical Chemistry Letters}\ }\textbf {\bibinfo {volume} {4}},\
  \bibinfo {pages} {132} (\bibinfo {year} {2013})}\BibitemShut {NoStop}%
\bibitem [{\citenamefont {Merlet}\ \emph {et~al.}(2013)\citenamefont {Merlet},
  \citenamefont {P{\'{e}}an}, \citenamefont {Rotenberg}, \citenamefont
  {Madden}, \citenamefont {Simon},\ and\ \citenamefont
  {Salanne}}]{Merlet2013SimulatingSurfaces}%
  \BibitemOpen
  \bibfield  {author} {\bibinfo {author} {\bibfnamefont {C.}~\bibnamefont
  {Merlet}}, \bibinfo {author} {\bibfnamefont {C.}~\bibnamefont {P{\'{e}}an}},
  \bibinfo {author} {\bibfnamefont {B.}~\bibnamefont {Rotenberg}}, \bibinfo
  {author} {\bibfnamefont {P.~A.}\ \bibnamefont {Madden}}, \bibinfo {author}
  {\bibfnamefont {P.}~\bibnamefont {Simon}}, \ and\ \bibinfo {author}
  {\bibfnamefont {M.}~\bibnamefont {Salanne}},\ }\href {\doibase
  10.1021/jz3019226} {\bibfield  {journal} {\bibinfo  {journal} {Journal of
  Physical Chemistry Letters}\ }\textbf {\bibinfo {volume} {4}},\ \bibinfo
  {pages} {264} (\bibinfo {year} {2013})}\BibitemShut {NoStop}%
\bibitem [{\citenamefont {Vatamanu}\ \emph {et~al.}(2017)\citenamefont
  {Vatamanu}, \citenamefont {Bedrov},\ and\ \citenamefont
  {Borodin}}]{Vatamanu2017OnLayers}%
  \BibitemOpen
  \bibfield  {author} {\bibinfo {author} {\bibfnamefont {J.}~\bibnamefont
  {Vatamanu}}, \bibinfo {author} {\bibfnamefont {D.}~\bibnamefont {Bedrov}}, \
  and\ \bibinfo {author} {\bibfnamefont {O.}~\bibnamefont {Borodin}},\ }\href
  {\doibase 10.1080/08927022.2017.1279287} {\bibfield  {journal} {\bibinfo
  {journal} {Molecular Simulation}\ }\textbf {\bibinfo {volume} {43}},\
  \bibinfo {pages} {838} (\bibinfo {year} {2017})}\BibitemShut {NoStop}%
\bibitem [{\citenamefont {Noh}\ and\ \citenamefont
  {Jung}(2019)}]{Noh2019UnderstandingSimulations}%
  \BibitemOpen
  \bibfield  {author} {\bibinfo {author} {\bibfnamefont {C.}~\bibnamefont
  {Noh}}\ and\ \bibinfo {author} {\bibfnamefont {Y.}~\bibnamefont {Jung}},\
  }\href {\doibase 10.1039/c8cp07200k} {\bibfield  {journal} {\bibinfo
  {journal} {Physical Chemistry Chemical Physics}\ }\textbf {\bibinfo {volume}
  {21}},\ \bibinfo {pages} {6790} (\bibinfo {year} {2019})}\BibitemShut
  {NoStop}%
\bibitem [{\citenamefont {Demir}\ and\ \citenamefont
  {Searles}(2020)}]{Demir2020InvestigationSimulations}%
  \BibitemOpen
  \bibfield  {author} {\bibinfo {author} {\bibfnamefont {B.}~\bibnamefont
  {Demir}}\ and\ \bibinfo {author} {\bibfnamefont {D.~J.}\ \bibnamefont
  {Searles}},\ }\href {\doibase 10.3390/nano10112181} {\bibfield  {journal}
  {\bibinfo  {journal} {Nanomaterials}\ }\textbf {\bibinfo {volume} {10}},\
  \bibinfo {pages} {2181} (\bibinfo {year} {2020})}\BibitemShut {NoStop}%
\bibitem [{\citenamefont {Seidl}\ \emph {et~al.}(2021)\citenamefont {Seidl},
  \citenamefont {H{\"{o}}rmann},\ and\ \citenamefont
  {Pastewka}}]{Seidl2021MolecularElectrolytes}%
  \BibitemOpen
  \bibfield  {author} {\bibinfo {author} {\bibfnamefont {C.}~\bibnamefont
  {Seidl}}, \bibinfo {author} {\bibfnamefont {J.~L.}\ \bibnamefont
  {H{\"{o}}rmann}}, \ and\ \bibinfo {author} {\bibfnamefont {L.}~\bibnamefont
  {Pastewka}},\ }\href {\doibase 10.1007/s11249-020-01395-6} {\bibfield
  {journal} {\bibinfo  {journal} {Tribology Letters}\ }\textbf {\bibinfo
  {volume} {69}},\ \bibinfo {pages} {1} (\bibinfo {year} {2021})}\BibitemShut
  {NoStop}%
\bibitem [{\citenamefont {Nakano}\ and\ \citenamefont
  {Sato}(2019)}]{Nakano2019ASimulations}%
  \BibitemOpen
  \bibfield  {author} {\bibinfo {author} {\bibfnamefont {H.}~\bibnamefont
  {Nakano}}\ and\ \bibinfo {author} {\bibfnamefont {H.}~\bibnamefont {Sato}},\
  }\href {\doibase 10.1063/1.5123365} {\bibfield  {journal} {\bibinfo
  {journal} {Journal of Chemical Physics}\ }\textbf {\bibinfo {volume} {151}},\
  \bibinfo {pages} {164123} (\bibinfo {year} {2019})}\BibitemShut {NoStop}%
\bibitem [{\citenamefont {Scalfi}\ \emph
  {et~al.}(2020{\natexlab{a}})\citenamefont {Scalfi}, \citenamefont {Dufils},
  \citenamefont {Reeves}, \citenamefont {Rotenberg},\ and\ \citenamefont
  {Salanne}}]{Scalfi2020ASimulations}%
  \BibitemOpen
  \bibfield  {author} {\bibinfo {author} {\bibfnamefont {L.}~\bibnamefont
  {Scalfi}}, \bibinfo {author} {\bibfnamefont {T.}~\bibnamefont {Dufils}},
  \bibinfo {author} {\bibfnamefont {K.~G.}\ \bibnamefont {Reeves}}, \bibinfo
  {author} {\bibfnamefont {B.}~\bibnamefont {Rotenberg}}, \ and\ \bibinfo
  {author} {\bibfnamefont {M.}~\bibnamefont {Salanne}},\ }\href {\doibase
  10.1063/5.0028232} {\bibfield  {journal} {\bibinfo  {journal} {Journal of
  Chemical Physics}\ }\textbf {\bibinfo {volume} {153}},\ \bibinfo {pages}
  {174704} (\bibinfo {year} {2020}{\natexlab{a}})}\BibitemShut {NoStop}%
\bibitem [{\citenamefont {Ahrens-Iwers}\ and\ \citenamefont
  {Mei{\ss}ner}(2021)}]{Ahrens-Iwers2021ConstantMesh}%
  \BibitemOpen
  \bibfield  {author} {\bibinfo {author} {\bibfnamefont {L.~J.}\ \bibnamefont
  {Ahrens-Iwers}}\ and\ \bibinfo {author} {\bibfnamefont {R.~H.}\ \bibnamefont
  {Mei{\ss}ner}},\ }\href {\doibase 10.1063/5.0063381} {\bibfield  {journal}
  {\bibinfo  {journal} {The Journal of Chemical Physics}\ }\textbf {\bibinfo
  {volume} {155}},\ \bibinfo {pages} {104104} (\bibinfo {year}
  {2021})}\BibitemShut {NoStop}%
\bibitem [{\citenamefont {Thompson}\ \emph {et~al.}(2022)\citenamefont
  {Thompson}, \citenamefont {Aktulga}, \citenamefont {Berger}, \citenamefont
  {Bolintineanu}, \citenamefont {Brown}, \citenamefont {Crozier}, \citenamefont
  {in~'t Veld}, \citenamefont {Kohlmeyer}, \citenamefont {Moore}, \citenamefont
  {Nguyen}, \citenamefont {Shan}, \citenamefont {Stevens}, \citenamefont
  {Tranchida}, \citenamefont {Trott},\ and\ \citenamefont {Plimpton}}]{LAMMPS}%
  \BibitemOpen
  \bibfield  {author} {\bibinfo {author} {\bibfnamefont {A.~P.}\ \bibnamefont
  {Thompson}}, \bibinfo {author} {\bibfnamefont {H.~M.}\ \bibnamefont
  {Aktulga}}, \bibinfo {author} {\bibfnamefont {R.}~\bibnamefont {Berger}},
  \bibinfo {author} {\bibfnamefont {D.~S.}\ \bibnamefont {Bolintineanu}},
  \bibinfo {author} {\bibfnamefont {W.~M.}\ \bibnamefont {Brown}}, \bibinfo
  {author} {\bibfnamefont {P.~S.}\ \bibnamefont {Crozier}}, \bibinfo {author}
  {\bibfnamefont {P.~J.}\ \bibnamefont {in~'t Veld}}, \bibinfo {author}
  {\bibfnamefont {A.}~\bibnamefont {Kohlmeyer}}, \bibinfo {author}
  {\bibfnamefont {S.~G.}\ \bibnamefont {Moore}}, \bibinfo {author}
  {\bibfnamefont {T.~D.}\ \bibnamefont {Nguyen}}, \bibinfo {author}
  {\bibfnamefont {R.}~\bibnamefont {Shan}}, \bibinfo {author} {\bibfnamefont
  {M.~J.}\ \bibnamefont {Stevens}}, \bibinfo {author} {\bibfnamefont
  {J.}~\bibnamefont {Tranchida}}, \bibinfo {author} {\bibfnamefont
  {C.}~\bibnamefont {Trott}}, \ and\ \bibinfo {author} {\bibfnamefont {S.~J.}\
  \bibnamefont {Plimpton}},\ }\href {\doibase 10.1016/j.cpc.2021.108171}
  {\bibfield  {journal} {\bibinfo  {journal} {Comp. Phys. Comm.}\ }\textbf
  {\bibinfo {volume} {271}},\ \bibinfo {pages} {108171} (\bibinfo {year}
  {2022})}\BibitemShut {NoStop}%
\bibitem [{\citenamefont {Tee}\ and\ \citenamefont
  {Bernhardt}(2022)}]{Tee2022Repo}%
  \BibitemOpen
  \bibfield  {author} {\bibinfo {author} {\bibfnamefont {S.}~\bibnamefont
  {Tee}}\ and\ \bibinfo {author} {\bibfnamefont {D.}~\bibnamefont
  {Bernhardt}},\ }\href@noop {} {\enquote {\bibinfo {title} {Source code for
  {USER-CONP2} add-on for {LAMMPS}},}\ } (\bibinfo {year} {2022}),\ \bibinfo
  {note} {{The} University of Queensland. Data Collection. DOI:
  \url{https://doi.org/10.48610/6b1122a}}\BibitemShut {NoStop}%
\bibitem [{Git(2021)}]{GitHubLink}%
  \BibitemOpen
  \href@noop {} {} (\bibinfo {year} {2021}),\ \bibinfo {note} {code hosted on
  GitHub at \url{https://github.com/srtee/lammps-USER-CONP2}}\BibitemShut
  {NoStop}%
\bibitem [{\citenamefont {Yeh}\ and\ \citenamefont
  {Berkowitz}(1999)}]{Yeh1999EwaldGeometry}%
  \BibitemOpen
  \bibfield  {author} {\bibinfo {author} {\bibfnamefont {I.~C.}\ \bibnamefont
  {Yeh}}\ and\ \bibinfo {author} {\bibfnamefont {M.~L.}\ \bibnamefont
  {Berkowitz}},\ }\href {\doibase 10.1063/1.479595} {\bibfield  {journal}
  {\bibinfo  {journal} {Journal of Chemical Physics}\ }\textbf {\bibinfo
  {volume} {111}},\ \bibinfo {pages} {3155} (\bibinfo {year}
  {1999})}\BibitemShut {NoStop}%
\bibitem [{\citenamefont {Dufils}\ \emph {et~al.}(2019)\citenamefont {Dufils},
  \citenamefont {Jeanmairet}, \citenamefont {Rotenberg}, \citenamefont
  {Sprik},\ and\ \citenamefont {Salanne}}]{Dufils2019SimulatingElectrode}%
  \BibitemOpen
  \bibfield  {author} {\bibinfo {author} {\bibfnamefont {T.}~\bibnamefont
  {Dufils}}, \bibinfo {author} {\bibfnamefont {G.}~\bibnamefont {Jeanmairet}},
  \bibinfo {author} {\bibfnamefont {B.}~\bibnamefont {Rotenberg}}, \bibinfo
  {author} {\bibfnamefont {M.}~\bibnamefont {Sprik}}, \ and\ \bibinfo {author}
  {\bibfnamefont {M.}~\bibnamefont {Salanne}},\ }\href {\doibase
  10.1103/PhysRevLett.123.195501} {\bibfield  {journal} {\bibinfo  {journal}
  {Physical Review Letters}\ }\textbf {\bibinfo {volume} {123}},\ \bibinfo
  {pages} {195501} (\bibinfo {year} {2019})}\BibitemShut {NoStop}%
\bibitem [{\citenamefont {Raiteri}\ \emph {et~al.}(2020)\citenamefont
  {Raiteri}, \citenamefont {Kraus},\ and\ \citenamefont
  {Gale}}]{Raiteri2020MolecularInterface}%
  \BibitemOpen
  \bibfield  {author} {\bibinfo {author} {\bibfnamefont {P.}~\bibnamefont
  {Raiteri}}, \bibinfo {author} {\bibfnamefont {P.}~\bibnamefont {Kraus}}, \
  and\ \bibinfo {author} {\bibfnamefont {J.~D.}\ \bibnamefont {Gale}},\ }\href
  {\doibase 10.1063/5.0027876} {\bibfield  {journal} {\bibinfo  {journal} {The
  Journal of Chemical Physics}\ }\textbf {\bibinfo {volume} {153}},\ \bibinfo
  {pages} {164714} (\bibinfo {year} {2020})}\BibitemShut {NoStop}%
\bibitem [{\citenamefont {Scalfi}\ \emph
  {et~al.}(2020{\natexlab{b}})\citenamefont {Scalfi}, \citenamefont {Limmer},
  \citenamefont {Coretti}, \citenamefont {Bonella}, \citenamefont {Madden},
  \citenamefont {Salanne},\ and\ \citenamefont
  {Rotenberg}}]{Scalfi2020ChargeEnsemble}%
  \BibitemOpen
  \bibfield  {author} {\bibinfo {author} {\bibfnamefont {L.}~\bibnamefont
  {Scalfi}}, \bibinfo {author} {\bibfnamefont {D.~T.}\ \bibnamefont {Limmer}},
  \bibinfo {author} {\bibfnamefont {A.}~\bibnamefont {Coretti}}, \bibinfo
  {author} {\bibfnamefont {S.}~\bibnamefont {Bonella}}, \bibinfo {author}
  {\bibfnamefont {P.~A.}\ \bibnamefont {Madden}}, \bibinfo {author}
  {\bibfnamefont {M.}~\bibnamefont {Salanne}}, \ and\ \bibinfo {author}
  {\bibfnamefont {B.}~\bibnamefont {Rotenberg}},\ }\href {\doibase
  10.1039/c9cp06285h} {\bibfield  {journal} {\bibinfo  {journal} {Physical
  Chemistry Chemical Physics}\ }\textbf {\bibinfo {volume} {22}},\ \bibinfo
  {pages} {10480} (\bibinfo {year} {2020}{\natexlab{b}})}\BibitemShut {NoStop}%
\bibitem [{\citenamefont {de~Leeuw}\ \emph {et~al.}(1980)\citenamefont
  {de~Leeuw}, \citenamefont {Perram},\ and\ \citenamefont
  {Smith}}]{deLeeuw1980SimulationConstants}%
  \BibitemOpen
  \bibfield  {author} {\bibinfo {author} {\bibfnamefont {S.~W.}\ \bibnamefont
  {de~Leeuw}}, \bibinfo {author} {\bibfnamefont {J.~W.}\ \bibnamefont
  {Perram}}, \ and\ \bibinfo {author} {\bibfnamefont {E.~R.}\ \bibnamefont
  {Smith}},\ }\href {\doibase 10.1098/rspa.1980.0135} {\bibfield  {journal}
  {\bibinfo  {journal} {Proceedings of the Royal Society of London. A.
  Mathematical and Physical Sciences}\ }\textbf {\bibinfo {volume} {373}},\
  \bibinfo {pages} {27} (\bibinfo {year} {1980})}\BibitemShut {NoStop}%
\bibitem [{\citenamefont {Allen}\ and\ \citenamefont
  {Tildesley}(1989)}]{Allen1989ComputerLiquids}%
  \BibitemOpen
  \bibfield  {author} {\bibinfo {author} {\bibfnamefont {M.~P.}\ \bibnamefont
  {Allen}}\ and\ \bibinfo {author} {\bibfnamefont {D.~J.}\ \bibnamefont
  {Tildesley}},\ }\href@noop {} {\emph {\bibinfo {title} {{Computer Simulation
  of Liquids}}}}\ (\bibinfo  {publisher} {Clarendon Press},\ \bibinfo {year}
  {1989})\BibitemShut {NoStop}%
\bibitem [{\citenamefont {Kawata}\ \emph {et~al.}(2001)\citenamefont {Kawata},
  \citenamefont {Mikami},\ and\ \citenamefont
  {Nagashima}}]{Kawata2001RapidPeriodicity}%
  \BibitemOpen
  \bibfield  {author} {\bibinfo {author} {\bibfnamefont {M.}~\bibnamefont
  {Kawata}}, \bibinfo {author} {\bibfnamefont {M.}~\bibnamefont {Mikami}}, \
  and\ \bibinfo {author} {\bibfnamefont {U.}~\bibnamefont {Nagashima}},\ }\href
  {\doibase 10.1063/1.1395564} {\bibfield  {journal} {\bibinfo  {journal}
  {Journal of Chemical Physics}\ }\textbf {\bibinfo {volume} {115}},\ \bibinfo
  {pages} {4457} (\bibinfo {year} {2001})}\BibitemShut {NoStop}%
\bibitem [{\citenamefont {Br{\'{o}}dka}\ and\ \citenamefont
  {Grzybowski}(2002)}]{Brodka2002ElectrostaticSummation}%
  \BibitemOpen
  \bibfield  {author} {\bibinfo {author} {\bibfnamefont {A.}~\bibnamefont
  {Br{\'{o}}dka}}\ and\ \bibinfo {author} {\bibfnamefont {A.}~\bibnamefont
  {Grzybowski}},\ }\href {\doibase 10.1063/1.1513151} {\bibfield  {journal}
  {\bibinfo  {journal} {Journal of Chemical Physics}\ }\textbf {\bibinfo
  {volume} {117}},\ \bibinfo {pages} {8208} (\bibinfo {year}
  {2002})}\BibitemShut {NoStop}%
\bibitem [{\citenamefont {Zhang}\ \emph {et~al.}(2020)\citenamefont {Zhang},
  \citenamefont {Sayer}, \citenamefont {Hutter},\ and\ \citenamefont
  {Sprik}}]{Zhang2020ModellingDynamics}%
  \BibitemOpen
  \bibfield  {author} {\bibinfo {author} {\bibfnamefont {C.}~\bibnamefont
  {Zhang}}, \bibinfo {author} {\bibfnamefont {T.}~\bibnamefont {Sayer}},
  \bibinfo {author} {\bibfnamefont {J.}~\bibnamefont {Hutter}}, \ and\ \bibinfo
  {author} {\bibfnamefont {M.}~\bibnamefont {Sprik}},\ }\href {\doibase
  10.1088/2515-7655/ab9d8c} {\bibfield  {journal} {\bibinfo  {journal} {Journal
  of Physics: Energy}\ }\textbf {\bibinfo {volume} {2}},\ \bibinfo {pages}
  {032005} (\bibinfo {year} {2020})}\BibitemShut {NoStop}%
\bibitem [{\citenamefont {Dufils}\ \emph {et~al.}(2021)\citenamefont {Dufils},
  \citenamefont {Sprik},\ and\ \citenamefont
  {Salanne}}]{Dufils2021ComputationalSimulations}%
  \BibitemOpen
  \bibfield  {author} {\bibinfo {author} {\bibfnamefont {T.}~\bibnamefont
  {Dufils}}, \bibinfo {author} {\bibfnamefont {M.}~\bibnamefont {Sprik}}, \
  and\ \bibinfo {author} {\bibfnamefont {M.}~\bibnamefont {Salanne}},\ }\href
  {\doibase 10.1021/acs.jpclett.1c01131} {\bibfield  {journal} {\bibinfo
  {journal} {The Journal of Physical Chemistry Letters}\ }\textbf {\bibinfo
  {volume} {12}},\ \bibinfo {pages} {4357} (\bibinfo {year}
  {2021})}\BibitemShut {NoStop}%
\bibitem [{\citenamefont {Croteau}\ \emph {et~al.}(2009)\citenamefont
  {Croteau}, \citenamefont {K.~Bertram},\ and\ \citenamefont
  {N.~Patey}}]{Croteau2009SimulationConditions}%
  \BibitemOpen
  \bibfield  {author} {\bibinfo {author} {\bibfnamefont {T.}~\bibnamefont
  {Croteau}}, \bibinfo {author} {\bibfnamefont {A.}~\bibnamefont {K.~Bertram}},
  \ and\ \bibinfo {author} {\bibfnamefont {G.}~\bibnamefont {N.~Patey}},\
  }\href {\doibase 10.1021/jp902453f} {\bibfield  {journal} {\bibinfo
  {journal} {The Journal of Physical Chemistry A}\ }\textbf {\bibinfo {volume}
  {113}},\ \bibinfo {pages} {7826} (\bibinfo {year} {2009})}\BibitemShut
  {NoStop}%
\bibitem [{\citenamefont {Ren}\ \emph {et~al.}(2020)\citenamefont {Ren},
  \citenamefont {K.~Bertram},\ and\ \citenamefont
  {N.~Patey}}]{Ren2020EffectsWater}%
  \BibitemOpen
  \bibfield  {author} {\bibinfo {author} {\bibfnamefont {Y.}~\bibnamefont
  {Ren}}, \bibinfo {author} {\bibfnamefont {A.}~\bibnamefont {K.~Bertram}}, \
  and\ \bibinfo {author} {\bibfnamefont {G.}~\bibnamefont {N.~Patey}},\ }\href
  {\doibase 10.1021/acs.jpcb.0c01695} {\bibfield  {journal} {\bibinfo
  {journal} {The Journal of Physical Chemistry B}\ }\textbf {\bibinfo {volume}
  {124}},\ \bibinfo {pages} {4605} (\bibinfo {year} {2020})}\BibitemShut
  {NoStop}%
\bibitem [{\citenamefont {Wang}\ \emph {et~al.}(2016)\citenamefont {Wang},
  \citenamefont {Olmsted}, \citenamefont {Asta},\ and\ \citenamefont
  {Laird}}]{Wang2016ElectricCapacitors}%
  \BibitemOpen
  \bibfield  {author} {\bibinfo {author} {\bibfnamefont {Z.}~\bibnamefont
  {Wang}}, \bibinfo {author} {\bibfnamefont {D.~L.}\ \bibnamefont {Olmsted}},
  \bibinfo {author} {\bibfnamefont {M.}~\bibnamefont {Asta}}, \ and\ \bibinfo
  {author} {\bibfnamefont {B.~B.}\ \bibnamefont {Laird}},\ }\href {\doibase
  10.1088/0953-8984/28/46/464006} {\bibfield  {journal} {\bibinfo  {journal}
  {Journal of Physics Condensed Matter}\ }\textbf {\bibinfo {volume} {28}},\
  \bibinfo {pages} {464006} (\bibinfo {year} {2016})}\BibitemShut {NoStop}%
\bibitem [{\citenamefont {Cole}\ and\ \citenamefont
  {Klein}(1983)}]{Cole1983TheGraphite}%
  \BibitemOpen
  \bibfield  {author} {\bibinfo {author} {\bibfnamefont {M.~W.}\ \bibnamefont
  {Cole}}\ and\ \bibinfo {author} {\bibfnamefont {J.~R.}\ \bibnamefont
  {Klein}},\ }\href {\doibase 10.1016/0039-6028(83)90808-7} {\bibfield
  {journal} {\bibinfo  {journal} {Surface Science}\ }\textbf {\bibinfo {volume}
  {124}},\ \bibinfo {pages} {547} (\bibinfo {year} {1983})}\BibitemShut
  {NoStop}%
\bibitem [{\citenamefont {Bazant}\ \emph {et~al.}(2009)\citenamefont {Bazant},
  \citenamefont {Kilic}, \citenamefont {Storey},\ and\ \citenamefont
  {Ajdari}}]{Bazant2009TowardsSolutions}%
  \BibitemOpen
  \bibfield  {author} {\bibinfo {author} {\bibfnamefont {M.~Z.}\ \bibnamefont
  {Bazant}}, \bibinfo {author} {\bibfnamefont {M.~S.}\ \bibnamefont {Kilic}},
  \bibinfo {author} {\bibfnamefont {B.~D.}\ \bibnamefont {Storey}}, \ and\
  \bibinfo {author} {\bibfnamefont {A.}~\bibnamefont {Ajdari}},\ }\href
  {\doibase 10.1016/j.cis.2009.10.001} {\bibfield  {journal} {\bibinfo
  {journal} {Advances in Colloid and Interface Science}\ }\textbf {\bibinfo
  {volume} {152}},\ \bibinfo {pages} {48} (\bibinfo {year} {2009})}\BibitemShut
  {NoStop}%
\bibitem [{\citenamefont {Merlet}\ \emph {et~al.}(2011)\citenamefont {Merlet},
  \citenamefont {Salanne}, \citenamefont {Rotenberg},\ and\ \citenamefont
  {Madden}}]{Merlet2011ImidazoliumSimulations}%
  \BibitemOpen
  \bibfield  {author} {\bibinfo {author} {\bibfnamefont {C.}~\bibnamefont
  {Merlet}}, \bibinfo {author} {\bibfnamefont {M.}~\bibnamefont {Salanne}},
  \bibinfo {author} {\bibfnamefont {B.}~\bibnamefont {Rotenberg}}, \ and\
  \bibinfo {author} {\bibfnamefont {P.~A.}\ \bibnamefont {Madden}},\ }\href
  {\doibase 10.1021/jp205461g} {\bibfield  {journal} {\bibinfo  {journal}
  {Journal of Physical Chemistry C}\ }\textbf {\bibinfo {volume} {115}},\
  \bibinfo {pages} {16613} (\bibinfo {year} {2011})}\BibitemShut {NoStop}%
\bibitem [{\citenamefont {Hockney}\ and\ \citenamefont
  {Eastwood}(1988)}]{Hockney1988ComputerParticles}%
  \BibitemOpen
  \bibfield  {author} {\bibinfo {author} {\bibfnamefont {R.~W.}\ \bibnamefont
  {Hockney}}\ and\ \bibinfo {author} {\bibfnamefont {J.~W.}\ \bibnamefont
  {Eastwood}},\ }\href@noop {} {\emph {\bibinfo {title} {{Computer Simulation
  Using Particles}}}}\ (\bibinfo  {publisher} {Taylor and Francis Inc.},\
  \bibinfo {year} {1988})\BibitemShut {NoStop}%
\bibitem [{\citenamefont {Nos{\'{e}}}(1984{\natexlab{a}})}]{Nose1984AEnsemble}%
  \BibitemOpen
  \bibfield  {author} {\bibinfo {author} {\bibfnamefont {S.}~\bibnamefont
  {Nos{\'{e}}}},\ }\href {\doibase 10.1080/00268978400101201} {\bibfield
  {journal} {\bibinfo  {journal} {Molecular Physics}\ }\textbf {\bibinfo
  {volume} {52}},\ \bibinfo {pages} {255} (\bibinfo {year}
  {1984}{\natexlab{a}})}\BibitemShut {NoStop}%
\bibitem [{\citenamefont {Nos{\'{e}}}(1984{\natexlab{b}})}]{Nose1984AMethods}%
  \BibitemOpen
  \bibfield  {author} {\bibinfo {author} {\bibfnamefont {S.}~\bibnamefont
  {Nos{\'{e}}}},\ }\href {\doibase 10.1063/1.447334} {\bibfield  {journal}
  {\bibinfo  {journal} {The Journal of Chemical Physics}\ }\textbf {\bibinfo
  {volume} {81}},\ \bibinfo {pages} {511} (\bibinfo {year}
  {1984}{\natexlab{b}})}\BibitemShut {NoStop}%
\bibitem [{\citenamefont {Hoover}(1985)}]{Hoover1985CanonicalDistributions}%
  \BibitemOpen
  \bibfield  {author} {\bibinfo {author} {\bibfnamefont {W.~G.}\ \bibnamefont
  {Hoover}},\ }\href {\doibase 10.1103/PhysRevA.31.1695} {\bibfield  {journal}
  {\bibinfo  {journal} {Physical Review A}\ }\textbf {\bibinfo {volume} {31}},\
  \bibinfo {pages} {1695} (\bibinfo {year} {1985})}\BibitemShut {NoStop}%
\bibitem [{\citenamefont {Ryckaert}\ \emph {et~al.}(1977)\citenamefont
  {Ryckaert}, \citenamefont {Ciccotti},\ and\ \citenamefont
  {Berendsen}}]{Ryckaert1977NumericalN-alkanes}%
  \BibitemOpen
  \bibfield  {author} {\bibinfo {author} {\bibfnamefont {J.~P.}\ \bibnamefont
  {Ryckaert}}, \bibinfo {author} {\bibfnamefont {G.}~\bibnamefont {Ciccotti}},
  \ and\ \bibinfo {author} {\bibfnamefont {H.~J.}\ \bibnamefont {Berendsen}},\
  }\href {\doibase 10.1016/0021-9991(77)90098-5} {\bibfield  {journal}
  {\bibinfo  {journal} {Journal of Computational Physics}\ }\textbf {\bibinfo
  {volume} {23}},\ \bibinfo {pages} {327} (\bibinfo {year} {1977})}\BibitemShut
  {NoStop}%
\bibitem [{\citenamefont {Shinoda}\ \emph {et~al.}(2004)\citenamefont
  {Shinoda}, \citenamefont {Shiga},\ and\ \citenamefont
  {Mikami}}]{Shinoda2004RapidStress}%
  \BibitemOpen
  \bibfield  {author} {\bibinfo {author} {\bibfnamefont {W.}~\bibnamefont
  {Shinoda}}, \bibinfo {author} {\bibfnamefont {M.}~\bibnamefont {Shiga}}, \
  and\ \bibinfo {author} {\bibfnamefont {M.}~\bibnamefont {Mikami}},\ }\href
  {\doibase 10.1103/PhysRevB.69.134103} {\bibfield  {journal} {\bibinfo
  {journal} {Physical Review B - Condensed Matter and Materials Physics}\
  }\textbf {\bibinfo {volume} {69}},\ \bibinfo {pages} {134103} (\bibinfo
  {year} {2004})}\BibitemShut {NoStop}%
\bibitem [{\citenamefont {Tu}\ \emph {et~al.}(2020)\citenamefont {Tu},
  \citenamefont {Delmerico},\ and\ \citenamefont
  {McDaniel}}]{Tu2020InnerDynamics}%
  \BibitemOpen
  \bibfield  {author} {\bibinfo {author} {\bibfnamefont {Y.~J.}\ \bibnamefont
  {Tu}}, \bibinfo {author} {\bibfnamefont {S.}~\bibnamefont {Delmerico}}, \
  and\ \bibinfo {author} {\bibfnamefont {J.~G.}\ \bibnamefont {McDaniel}},\
  }\href {\doibase 10.1021/acs.jpcc.0c00299} {\bibfield  {journal} {\bibinfo
  {journal} {Journal of Physical Chemistry C}\ }\textbf {\bibinfo {volume}
  {124}},\ \bibinfo {pages} {2907} (\bibinfo {year} {2020})}\BibitemShut
  {NoStop}%
\bibitem [{\citenamefont {Ballenegger}\ \emph {et~al.}(2009)\citenamefont
  {Ballenegger}, \citenamefont {Arnold},\ and\ \citenamefont
  {Cerd{\`{a}}}}]{Ballenegger2009SimulationsConditions}%
  \BibitemOpen
  \bibfield  {author} {\bibinfo {author} {\bibfnamefont {V.}~\bibnamefont
  {Ballenegger}}, \bibinfo {author} {\bibfnamefont {A.}~\bibnamefont {Arnold}},
  \ and\ \bibinfo {author} {\bibfnamefont {J.~J.}\ \bibnamefont
  {Cerd{\`{a}}}},\ }\href {\doibase 10.1063/1.3216473} {\bibfield  {journal}
  {\bibinfo  {journal} {Journal of Chemical Physics}\ }\textbf {\bibinfo
  {volume} {131}},\ \bibinfo {pages} {094107} (\bibinfo {year}
  {2009})}\BibitemShut {NoStop}%
\end{thebibliography}%


%

\end{document}